\documentclass[%
 11pt,
 reprint,
 superscriptaddress,
 onecolumn,
 showkeys,
 amsmath,amssymb,
 aps,
]{revtex4}
\usepackage{setspace}
\usepackage[colorlinks, citecolor=red]{hyperref}
\usepackage{graphicx}
\usepackage{dcolumn}
\usepackage{bm}

\usepackage{listings}
\usepackage{color}
\usepackage{subfigure}
\usepackage{algorithm}
\usepackage{algorithmic}
\usepackage{bbm}
\usepackage{ntheorem}

\newtheorem{Theo}{Theorem}
\newtheorem*{Prf}{Proof}
\newtheorem{lemma}{Lemma}

\begin{document}


\title{Quantum homotopy perturbation method for nonlinear dissipative ordinary differential equations}
\author{Cheng Xue}
 \affiliation{CAS Key Laboratory of Quantum Information, University of Science
 and Technology of China, Hefei, Anhui, 230026, P. R. China}
 \affiliation{CAS Center For Excellence in Quantum Information and Quantum Physics, University of Science and Technology of China, Hefei, Anhui, 230026, P. R. China}
\author{Yu-Chun Wu}
 \email{wuyuchun@ustc.edu.cn} 
 \affiliation{CAS Key Laboratory of Quantum Information, University of Science
 and Technology of China, Hefei, Anhui, 230026, P. R. China}
 \affiliation{CAS Center For Excellence in Quantum Information and Quantum Physics, University of Science and Technology of China, Hefei, Anhui, 230026, P. R. China}
 \affiliation{Institute of Artificial Intelligence, Hefei Comprehensive National Science Center, Hefei, Anhui, 230026, P. R. China}
\author{Guo-Ping Guo}
 \email{gpguo@ustc.edu.cn} 
 \affiliation{CAS Key Laboratory of Quantum Information, University of Science
 and Technology of China, Hefei, Anhui, 230026, P. R. China}
 \affiliation{CAS Center For Excellence in Quantum Information and Quantum Physics, University of Science and Technology of China, Hefei, Anhui, 230026, P. R. China}
 \affiliation{Institute of Artificial Intelligence, Hefei Comprehensive National Science Center, Hefei, Anhui, 230026, P. R. China}
 \affiliation{Origin Quantum Computing Company Limited, Hefei, Anhui, 230026, P. R. China}

\begin{abstract}

  While quantum computing provides an exponential advantage in solving linear differential equations, there are relatively few quantum algorithms for solving nonlinear differential equations. 
  In our work, based on the homotopy perturbation method,
  we propose a quantum algorithm for solving $n$-dimensional nonlinear dissipative ordinary differential equations (ODEs). Our algorithm first converts the original nonlinear ODEs into the other nonlinear ODEs which can be embedded into finite-dimensional linear ODEs. Then we solve the embedded linear ODEs with quantum linear ODEs algorithm and obtain a state $\epsilon$-close to the normalized exact solution of the original nonlinear ODEs with success probability $\Omega(1)$. 
  The complexity of our algorithm is $O(g\eta T{\rm poly}(\log(nT/\epsilon)))$, where $\eta$, $g$ measure the decay of the solution. Our algorithm provides exponential improvement over the best classical algorithms or previous quantum algorithms in $n$ or $\epsilon$.

\end{abstract}

\maketitle


\section{Introduction}

Nonlinear differential equations appear in many fields, such as fluid dynamics, biology, finance, etc. In general, the analytical solutions of nonlinear differential equations cannot be obtained effectively. Numerical methods are
often used to solve nonlinear differential equations. However, when solving high-dimensional nonlinear differential equations, too many computing resources are required, which may exceed the computing power of classic computers. 
It is important to develop more efficient algorithms for solving nonlinear differential equations.

Quantum computing provides a promising way to speed up the solution of various equations. In recent years many quantum algorithms have been developed to solve various equations, such as system of linear equations\cite{harrow2009quantum,childs2017quantum,subacsi2019quantum, clader2013preconditioned}, Poisson equation\cite{cao2013quantum}, Dirac equation\cite{fillion2017algorithm}, heat equation\cite{linden2020quantum}, linear ordinary differential equations (ODEs)\cite{berry2014high,berry2017quantum,xin2020quantum,childs2020quantum}, linear partial ODEs\cite{arrazola2019quantum,childs2020high} and so on\cite{montanaro2016quantum,costa2019quantum,engel2019quantum}.

However, because of the linearity of quantum mechanics, solving nonlinear equations with quantum computing is challenging, some related algorithms are proposed\cite{leyton2008quantum,qian2019quantum,lubasch2020variational,liu2021efficient,lloyd2020quantum,budinski2021quantum,chen2021quantum,xue2021quantum,kyriienko2021solving}.
An early quantum algorithm for solving nonlinear ordinary differential equations (ODEs) is proposed in \cite{leyton2008quantum}, but the complexity of the algorithm increases exponentially with the evolution time.
In \cite{lubasch2020variational}, the authors proposed a variational quantum algorithm for solving nonlinear differential equations and demonstrate the algorithm by solving 1-dimensional nonlinear Schrodinger equation. 
However, when the equations become complicated, whether the parameterized quantum circuit used in their work is capable of expressing the solution of the problem and the optimization problem of the parameterized quantum circuit has not yet been concluded.
In \cite{liu2021efficient}, a quantum algorithm for solving nonlinear dissipative ODEs is constructed based on Carleman linearization\cite{carleman1932application,kowalski1991nonlinear}. This approach embeds the nonlinear ODEs into linear ODEs and solves the linear ODEs with quantum algorithm. The complexity of their algorithm is $O(\frac{T^2q}{\epsilon}{\rm poly}(\log(nT/\epsilon)))$ and $q$ measures decay of the solution.
In \cite{lloyd2020quantum}, nonlinear ODEs are embedded in Hilbert space, and the evolution of nonlinear ODEs is approximated by the evolution of subsystems in large systems.

Homotopy perturbation method\cite{he1999homotopy,babolian2009some,chakraverty2019advanced} is a semi-analytical technique for
solving linear as well as nonlinear ordinary/partial differential equations. 
This method, which is a combination of homotopy in topology and classic perturbation techniques, provides us with a convenient way to obtain analytic or approximate solutions for a wide variety of problems arising in different fields, such as Duffing equation\cite{he2003homotopy}, nonlinear wave equations\cite{he2005application} and so on.

In our work, we propose a quantum algorithm for solving time-independent quadratic nonlinear dissipative ODEs. The more general nonlinear ODEs can be reduced to the quadratic ODEs by introducing additional variables\cite{kerner1981universal,forets2017explicit}.
Our algorithm uses the homotopy perturbation method to transform the problem into a series of nonlinear ODEs which have a special structure. The transformed nonlinear ODEs can be embedded into linear ODEs with a technique similar to Carleman linearization. Then the linear ODEs are solved with quantum linear ODEs algorithm proposed in \cite{berry2017quantum}. Finally, we measure some qubit registers and obtain a state $\epsilon$-close to the normalized exact solution $u(T)/\Vert u(T)\Vert$ at evolution time $T$.

Our work is similar with Liu's work\cite{liu2021efficient}, here we list some differences: (1) The truncation method is different, our work uses homotopy perturbation method and \cite{liu2021efficient} uses Carleman linearization. The convergence condition of homotopy perturbation method and Carleman linearization are different. In \cite{liu2021efficient}, Liu et al. define $R=\frac{\Vert u_{in}\Vert \Vert F_2\Vert}{\vert Re(\lambda_1)\vert}$(Here we omit the inhomogeneity term $F_0$ in their definition) and the convergence condition is $R<1$. In our work, we define $K=\frac{4\Vert u_{in}\Vert \Vert F_2\Vert}{\vert Re(\lambda_1)\vert}=4R$, the convergence condition is $K<1$, the factor '$4$' is caused by the homotopy perturbation method. By Corollary $\bm 1$ in \cite{liu2021efficient} and Lemma \ref{theo-error-1}, the truncation errors of Carleman linearization and homotopy perturbation method decrease exponentially with the truncation order $c$ as $\Vert u_{in}\Vert R^c(1-e^{Re(\lambda_1)t})^c$ and $\frac{K^{c+2}}{1-K}$, respectively. (2)The dependence of our algorithm on the error $\epsilon$ and evolution time $T$ in our algorithm is $O(T\rm{poly}(\log(1/\epsilon)))$, which is $O(T^2/\epsilon)$ in \cite{liu2021efficient}. Therefore the complexity of our algorithm is exponentially improved on $\epsilon$ compared to \cite{liu2021efficient}, the cost of this improvement is a stronger constraint on the problem, i.e., $K<\sqrt{2}/2$ and $\frac{(c+1)\vert Re(\lambda_1)\vert }{\Vert F_2 \Vert}\leq 1$.

This paper is organized as follows. 
Sect.\ref{sec-problem} introduces the quadratic ODEs to be solved. Then we show the details of quantum homotopy perturbation method in Sect.\ref{sec-homotopy}. 
Initial state preparation and oracle construction of matrix $A$ are discussed in Sect.\ref{sec-state-oracle}. 
In the following three sections we analyze our method from different aspects: 
Sect.\ref{sec-condition-num} gives an upper bound of the condition number of the linear system to be solved. 
Sect.\ref{sec-solution-error} analyzes the solution error of our method. 
Sect.\ref{sec-success-prob} gives a lower bound of success probability. 
Next, the main result of our work is proved in Sect.\ref{sec-main-result}. 
Finally, we conclude our work with a discussion of the result and some open problems in Sect.\ref{sec-discussion}. 

\section{Quadratic ODEs}\label{sec-problem}

We focus on an initial value problem described by the $n$-dimensional quadratic ODEs.
The problem to be solved is defined as
\begin{equation}\label{nonlinear_eq}
  \frac{du}{dt}=F_1u+F_2u^{\otimes 2},\ u(0)=u_{in},
\end{equation}
where $u$, $u_{in}\in \mathbbm{R}^{n}$, $F_1\in \mathbbm{R}^{n\times n}$, $F_2\in \mathbbm{R}^{n\times n^2}$ are time independent sparse matrices. The sparsity of $F_1$, $F_2$ is $s$, which means the number of non-zero elements in each row or column of $F_1,F_2$ does not exceed $s$. 
We assume $F_1$ is a normal matrix and the eigenvalues $\lambda_i$ of $F_1$ satisfy $Re(\lambda_{n}) \leq\dots\leq Re(\lambda_{1})<0$.
We also assume oracles $O_{F1}$, $O_{F2}$ and $O_u$ are given, $O_{F1}$, $O_{F2}$ are used to extract the non-zero position and value of $F_1$, $F_2$ respectively, and $|u_{in}\rangle$ is prepared with $O_u$. In specific, $O_{F1}$, $O_{F2}$ and $O_u$ are defined as
\begin{equation}\label{f1f2oracle}
\begin{aligned}
    &O_{F11}|j\rangle|k\rangle=|j\rangle|f_1(j,k)\rangle, & 0\leq j\leq n-1,\ 0\leq k\leq s-1,\\
    &O_{F12}|j\rangle|k\rangle|z\rangle=|j\rangle|k\rangle|z\oplus {F_1}_{j,k}\rangle, & 0\leq j,\ k\leq n-1,\\
    &O_{F13}|j\rangle|0\rangle=|j\rangle|g(j)\rangle, & 0\leq j\leq n-1,\\
    &O_{F21}|j\rangle|k\rangle=|j\rangle|f_2(j,k)\rangle, & 0\leq j\leq n-1,\ 0\leq k\leq s-1,\\
    &O_{F22}|j\rangle|k\rangle|z\rangle=|j\rangle|k\rangle|z\oplus {F_2}_{j,k}\rangle, & 0\leq j\leq n-1,\ 0\leq k\leq n^2-1,\\
    &O_u|0\rangle=|u_{in}/\Vert u_{in}\Vert \rangle,
\end{aligned}
\end{equation}
where $f_1(j,k)$ and $f_2(j,k)$ represent the column number of the $k$-th non-zero element in $j$-th row of $F_1$, $F_2$ respectively, $g(j)$ satisfies $f_{1}(j,g(j))=j$, here we treat the diagonal element of $F_1$ as a non-zero element. $O_{F13}$ is used to construct an oracle of a matrix related to $F_1$, the details are shown in the proof of Lemma \ref{lemma-bm-new}. 

We define a parameter $K$ which characterizes the nonlinearity of Eq.(\ref{nonlinear_eq}), 
\begin{equation}\label{parameter-r-new}
    K:=\frac{4\Vert u_{in} \Vert \Vert F_2 \Vert}{\vert Re(\lambda_1) \vert}.
\end{equation}
We assume $K\geq \Vert u_{in}\Vert$, if this is not satisfied, we rescale $u$ to $ \zeta u$ with a suitable constant $\zeta$ which keeps $\frac{4\Vert u_{in} \Vert \Vert F_2 \Vert}{\vert Re(\lambda_1) \vert}$ unchanged and makes $K\geq \Vert u_{in}\Vert$. In this paper we use spectral norm, it means $\Vert \cdot \Vert=\Vert \cdot \Vert_2$.

\section{Quantum Homotopy Perturbation Method}\label{sec-homotopy}

In this section, we give the whole process of quantum homotopy perturbation method for solving Eq.(\ref{nonlinear_eq}). It contains four steps: 
\begin{itemize}
    \item [(1)]Using homotopy perturbation method to transform Eq.(\ref{nonlinear_eq}) into Eq.(\ref{linear_differential}), a series of nonlinear ODEs with variable $\nu_i$.
    \item [(2)]Embedding Eq.(\ref{linear_differential}) into Eq.(\ref{eq-linear}), linear ODEs with variable $\vec{y}$.
    \item [(3)]Solving Eq.(\ref{eq-linear}) with quantum algorithm proposed in \cite{berry2017quantum}. 
    \item [(4)]Measurement. 
\end{itemize}
The following four subsections introduce the details of these four steps.

\subsection{Homotopy perturbation method}\label{3_2}

Firstly, we introduce the process of homotopy perturbation method\cite{he1999homotopy,babolian2009some,chakraverty2019advanced} for solving Eq.(\ref{nonlinear_eq}). Using homotopy method, we construct homotopy $\nu(t,p):\mathbbm{R^+}\times [0,1]\to \mathbbm{R}^{n}$, which satisfies
\begin{equation}\label{homotopy_method}
  H(\nu,p)=\frac{d\nu}{dt}-F_1\nu-pF_2\nu^{\otimes 2}=0,\ \nu(0,p)=u_{in}.
\end{equation}
Assuming $\nu$ is represented as
\begin{equation}\label{nu_exp}
  \nu=\nu_0+p\nu_1+p^2\nu_2+\dots+p^c\nu_c.
\end{equation}
Substituting Eq.(\ref{nu_exp}) into Eq.(\ref{homotopy_method}), then equating the terms with identical powers of $p$, we have the following equations:
\begin{align}\label{linear_differential}
  &\frac{d\nu_0}{dt}-F_1\nu_0=0,\ \nu_0(0)=u_{in},\notag\\
  &\frac{d\nu_1}{dt}-F_1\nu_1-F_2\nu_0\otimes\nu_0=0,\ \nu_1(0)=0,\notag\\
  &\frac{d\nu_2}{dt}-F_1\nu_2-F_2(\nu_0\otimes\nu_1+\nu_1\otimes\nu_0)=0,\ \nu_2(0)=0,\notag\\
  &\dots\notag\\
  &\frac{d\nu_c}{dt}-F_1\nu_c-F_2\sum_{j=0}^{c-1}{\nu_{j}\otimes\nu_{c-1-j}}=0,\ \nu_c(0)=0.
\end{align}
When $p=1$ in Eq.(\ref{homotopy_method}), we have 
\begin{equation}\label{exp_solution}
  \tilde{u}=\nu_0+\nu_1+\nu_2+\dots+\nu_c.
\end{equation}
The difference between $\tilde{u}$ and $u$ is analyzed in Sect.\ref{subsec-hpm-error}.

\subsection{Linear embedding}\label{sec-combination}

Secondly, Eq.(\ref{linear_differential}) is embedded into the linear ODEs defined in Eq.(\ref{eq-linear}):
\begin{equation}\label{eq-linear}
    \frac{d\vec{y}}{dt}=A\vec{y},\ \vec{y}(0)=y_{in},
\end{equation}
where $\vec{y}=[\vec{y}_0,\vec{y}_1,\dots,\vec{y}_c]$, $\vec{y}_i$ satisfies
\begin{equation}
    \vec{y}_i=\left\{\begin{array}{ll}
        [\nu_0+\nu_1+\dots+\nu_c], & i=0, \\
        \left[\vec{y}_{i,0},\vec{y}_{i,1},\dots,\vec{y}_{i,\beta_i-1}\right], & 1\leq i \leq c,
        \end{array}\right.
\end{equation}
where $\beta_i$ represents the number of items in $\vec{y}_i$, $\vec{y}_{i,j}$ represents $j$-th item of $\vec{y}_i$, it is represented as $\vec{y}_{i,j}=\otimes_{k=0}^{i}{\nu_{a_{i,j,k}}}$,
$a_{i,j,k}$ satisfies
\begin{equation}\label{eq-aijk}
    a_{i,j,k}\geq 0,\  i+1\leq\sum_{k=0}^{i}{(a_{i,j,k}+1)}\leq c+1.
\end{equation}
By Eq.(\ref{eq-aijk}), $\beta_i$ satisfies
\begin{equation}
    \beta_i=\left\{\begin{array}{ll}
        1, & i=0, \\
        \sum_{k=i}^{c}{\binom{k}{i}}, & 1\leq i \leq c.
        \end{array}\right.  
\end{equation}
We set $\vec{y}_{i,0}=\nu_0^{\otimes {i+1}},\ i=1,2,\dots,c$, then by Eq.(\ref{linear_differential}),
$y_{in}$ is written as
\begin{equation}\label{eq-x-in}
    y_{in}=[[u_{in}],[u_{in}^{\otimes 2},0,\dots,0],[u_{in}^{\otimes 3},0,\dots,0]\dots,[u_{in}^{\otimes c+1}]].
\end{equation}
We define $\vec{a}_{i,j}=[a_{i,j,0},a_{i,j,1},\dots,a_{i,j,i}]$, the mapping $\vec{a}_{i,j} \mapsto j$ is one-to-one mapping, so we can construct the following two operations
\begin{equation}
    O_{a1}|\vec{a}_{i,j}\rangle|0\rangle=|\vec{a}_{i,j}\rangle|j\rangle,
\end{equation}
\begin{equation}
    O_{a2}|i\rangle|j\rangle|0\rangle=|i\rangle|j\rangle|\vec{a}_{i,j}\rangle
\end{equation}
with $O(c)$-qubit quantum arithmetic circuit, and the gate complexity is $O({\rm poly}(c))$\cite{nielsen2002quantum}. $O({\rm poly}(c))$ will not influence the complexity of our algorithm, so the complexity of $O_{a1}$ and $O_{a2}$ can be ignored in the following analysis.
The dimension of $\vec{y}_i$ is $n^{i+1}\beta_i$, so the dimension of $\vec{y}$ is 
\begin{equation}
    N=\sum_{i=0}^{c}{n^{i+1}\beta_i}=(n+1)^{c+1}-1-cn\approx (n+1)^{c+1}.
\end{equation}

Next we analyze the structure of matrix $A$. We have 
\begin{align}\label{eq-0517-v1}
    &\frac{d \vec{y}_{i,j}}{d t}=(\sum_{j=0}^{i}{I_n^{\otimes j}\otimes F_1 \otimes I_n^{\otimes i-j}})\vec{y}_{i,j}\notag\\
    &+\sum_{k=0}^{i}{(I_n^k\otimes F_2\otimes I_n^{\otimes {i-k}})\nu_{a_{i,j,0}}\otimes\dots\otimes \nu_{a_{i,j,k-1}}\otimes(\sum_{l=0}^{a_{i,j,k}-1}{\nu_l\otimes\nu_{a_{i,j,k}-1-l}})\otimes\nu_{a_{i,j,k+1}}\otimes\dots \otimes \nu_{a_{i,j,i}}},
\end{align}
where $\nu_{a_{i,j,0}}\otimes\dots\otimes \nu_{a_{i,j,k-1}}\otimes \nu_l\otimes\nu_{a_{i,j,k}-1-l}\otimes\nu_{a_{i,j,k+1}}\otimes \dots\otimes \nu_{a_{i,j,i}}\in \vec{y}_{i+1}$, so Eq.(\ref{eq-linear}) can be written as
\begin{equation}
    \frac{\mathrm{d}}{\mathrm{d} t}\left(\begin{array}{c}
    \vec{y}_{0} \\
    \vec{y}_{1} \\
    \vdots \\
    \vec{y}_{c-1} \\
    \vec{y}_{c}
    \end{array}\right)=\left(\begin{array}{cccccc}
    A_{0,0} &A_{0,1} & & & \\
    & A_{1,1} &A_{1,2} & &\\
    & &\ddots &\ddots  & \\
    & & &A_{c-1,c-1} &A_{c-1,c}  \\
    & & & &A_{c,c}
    \end{array}\right)\left(\begin{array}{c}
        \vec{y}_{0} \\
        \vec{y}_{1} \\
        \vdots \\
        \vec{y}_{c-1} \\
        \vec{y}_{c}
        \end{array}\right),
\end{equation}
$A_{i,i}$ is $n^{i+1}\beta_i$ dimensional square matrix, which is represented as
\begin{equation}\label{eq-aii-def}
    A_{i,i}=I_{\beta_i}\otimes(\sum_{j=0}^{i}{I_n^j\otimes F_1 \otimes I_n^{\otimes i-j}}),
\end{equation}
$A_{i,i+1}$ is $n^{i+1}\beta_i\times n^{i+2}\beta_{i+1}$ dimensional matrix, its elements are determined by Eq.(\ref{eq-0517-v1}). 
$|y(t)\rangle$ is defined to represent $\vec{y}(t)$:
\begin{equation}\label{def-vec-x}
    |y(t)\rangle=\sum_{i=0}^{c}{\sum_{j=0}^{\beta_i-1}{|i,j\rangle|y_{i,j}(t)\rangle}}.
\end{equation}

\subsection{Quantum linear ODEs algorithm}\label{sec-qldea}

Thirdly, Eq.(\ref{eq-linear}) is solved with the quantum algorithm proposed in \cite{berry2017quantum}.
$\vec{y}(t)$ is written as
\begin{equation}
    \vec{y}(t)=e^{At}\vec{y}(0).
\end{equation}
We define $T_k(z):=\sum_{j=0}^{k}{\frac{z^j}{j!}}$.
When $k$ is large enough and the evolution time $h$ is relatively short (for example, $h\leq 1/\Vert A\Vert$), we have $\vec{y}(h)\approx T_k(Ah)\vec{y}(0)$.
This approximate solution can be used as the initial condition for the next approximation, repeating this procedure $m$ steps we have the approximation of $\vec{y}(mh)$. 

Next we introduce the details of the algorithm proposed in \cite{berry2017quantum}.
Let $m,k,p\in \mathbbm{Z}^{+}$ and define
\begin{equation}\label{eq-cmkp}
    \begin{aligned}
    C_{m, k, p}(A):=& \sum_{j=0}^{d}|j\rangle\langle j|\otimes I-\sum_{i=0}^{m-1} \sum_{j=1}^{k}| i(k+1)+j\rangle\langle i(k+1)+j-1| \otimes A / j \\
    &-\sum_{i=0}^{m-1} \sum_{j=0}^{k}|(i+1)(k+1)\rangle\langle i(k+1)+j|\otimes I-\sum_{j=d-p+1}^{d}| j\rangle\langle j-1| \otimes I,
    \end{aligned}
\end{equation}
where $d:=m(k+1)+p$, $I$ is an $N$-dimensional unit matrix. We consider the linear system
\begin{equation}\label{3-30-2}
    C_{m, k, p}(Ah)|x\rangle=|0\rangle|y_{in}\rangle,
\end{equation}
where $|y_{in}\rangle \in\mathbbm{C}^N,h\in\mathbbm{R}^+$. After evolving $m$ steps, the approximate solution of $k$-order Taylor series is obtained, and the solution remains unchanged at $p$ steps.
The solution of Eq.(\ref{3-30-2}) is represented as $|x\rangle=C_{m, k, p}(A h)^{-1}|0\rangle\left|y_{in}\right\rangle$,
it can also be written as
\begin{equation}\label{eq-x-sol}
    |x\rangle=\sum_{i=0}^{m-1} \sum_{j=0}^{k}|i(k+1)+j\rangle\left|x_{i, j}\right\rangle+\sum_{j=0}^{p}|m(k+1)+j\rangle\left|x_{m, j}\right\rangle.
\end{equation}
By Eq.(\ref{eq-cmkp}), $|x_{i,j}\rangle$ satisfies
\begin{equation}
    \begin{array}{l}
    |x_{0,0}\rangle=|y_{in}\rangle, \\
    |x_{i, 0}\rangle=\sum_{j=0}^{k}|x_{i-1, j}\rangle, \quad 1 \leq i \leq m, \\
    |x_{i, 1}\rangle  =Ah|x_{i, 0}\rangle,\quad 0 \leq i<m, \\
    |x_{i, j}\rangle=(A h / j)|x_{i, j-1}\rangle,\quad 0 \leq i<m, 2 \leq j \leq k, \\
    |x_{m, j}\rangle =|x_{m, j-1}\rangle,\quad 1 \leq j \leq p.
    \end{array}
\end{equation}
Then we have
\begin{equation}
    \begin{aligned}
    \left|x_{0,0}\right\rangle &=\left|y_{in}\right\rangle, \\
    \left|x_{0, j}\right\rangle &=\left((A h)^{j} / j !\right)\left|x_{0,0}\right\rangle,\quad 1\leq j\leq k, \\
    \left|x_{1,0}\right\rangle &=T_{k}(A h)\left|x_{0,0}\right\rangle  \approx \exp (A h)\left|y_{in}\right\rangle, \\
    \left|x_{1, j}\right\rangle &=\left((A h)^{j} / j !\right)\left|x_{1,0}\right\rangle,\quad 1\leq j\leq k, \\
    \left|x_{2,0}\right\rangle &=T_{k}(A h)\left|x_{1,0}\right\rangle \approx \exp (2 A h)\left|y_{in}\right\rangle, \\
    & \vdots \\
    \left|x_{m-1,0}\right\rangle &=T_{k}(A h)\left|x_{m-2,0}\right\rangle \approx \exp (A h(m-1))\left|y_{in}\right\rangle, \\
    \left|x_{m-1, j}\right\rangle &=\left((A h)^{j} / j !\right)\left|x_{m-1,0}\right\rangle,\quad 1\leq j\leq k, \\
    \left|x_{m, 0}\right\rangle &=T_{k}(A h)\left|x_{m-1,0}\right\rangle \approx \exp (A h m)\left|y_{in}\right\rangle, \\
    \left|x_{m, j}\right\rangle &=\left|x_{m, 0}\right\rangle  \approx \exp (A h m)\left|y_{in}\right\rangle,\quad 1\leq j\leq p,
    \end{aligned}
\end{equation}
$|x_{i,0}\rangle$ is the approximate solution of the system at time $ih$, $i=\in\{0,1,2,\dots,m\}$. $|x_{m,0}\rangle=|x_{m,1}\rangle=\dots=|x_{m,p}\rangle$ is the approximate solution of $\vec{y}(t)=e^{At}y_{in}$ at $t=mh$.

\subsection{Measurement}\label{sec-measurement}

Finally, we measure some qubit registers of $|x\rangle$ and get a state $\epsilon$-close to the normalized solution of Eq.(\ref{nonlinear_eq}). The measurement is divided into two steps: (1) Measure the first qubit register of $|x\rangle$ which is defined in Eq.(\ref{eq-x-sol}), if the result is $|m(k+1)+j\rangle,j=0,1,\dots,p$, we have $|y(t)\rangle$ in the second qubit register of $|x\rangle$. (2) Measure the first qubit register of $|y(t)\rangle$ which is defined in Eq.(\ref{def-vec-x}), if the result is $|0,0\rangle$, we get a state $\epsilon$-close to $|u(t)/\Vert u(t)\Vert\rangle$ in the qubit second register of $|y(t)\rangle$.

This measurement step is probabilistic, the success probability is analyzed in Sect.\ref{sec-success-prob}.

\section{State Preparation and Oracle Construction}\label{sec-state-oracle}

In this section, we give the preparation of $|y_{in}\rangle$ and oracle construction of $A$.

\subsection{State preparation}
We first discuss the preparation of $|y_{in}\rangle$, the result is shown in Lemma \ref{init-state}.
\begin{lemma}\label{init-state}
    By Eq.(\ref{eq-x-in}), $|y_{in}\rangle$ is defined as
    \begin{equation}
        |y_{in}\rangle=\frac{1}{\sqrt{M}}\sum_{i=0}^{c}{|i,0\rangle|u_{in}^{\otimes i+1}\rangle},
    \end{equation}
    where $M=\sum_{j=i}^{c}{\Vert u_{in}\Vert^{2(i+1)}}$.
    Given $O_u$ defined in Eq.(\ref{f1f2oracle}), $|y_{in}\rangle$ can be prepared by querying $O_u$ $O(c)$ times.
\end{lemma}
\begin{Prf}
    First we prepare
    \begin{equation}
        |\psi\rangle=\frac{1}{\sqrt{M}}\sum_{i=0}^{c}{\Vert u_{in}\Vert^{2(i+1)}|i,0\rangle}.
    \end{equation}
    Then we execute controlled $O_u$ operation $C-O_u=\sum_{i=0}^{c}{|i,0\rangle\langle i,0|\otimes O_u^{\otimes i+1}}$ on $|\psi\rangle$,
    \begin{equation}
        |y_{in}\rangle=\frac{1}{\sqrt{M}}\sum_{i=0}^{c}{|i,0\rangle|u_{in}^{\otimes i+1}\rangle}.
    \end{equation}
    The query complexity of $O_u$ is $O(c)$.
\end{Prf}

\subsection{Oracle construction of $A$}

Before introducing oracle construction of $A$,
we analyze some features of $A$, including sparsity, upper bound of $\Vert A \Vert$ and eigenvalue of $A$. The results are shown in Lemma \ref{lemma-4-1}, Lemma \ref{lemma-4-2} and Lemma \ref{lemma-4-3}.

\begin{lemma}\label{lemma-4-1}
    The sparsity of the matrix $A$ is $O(sc^2)$.
\end{lemma}
\begin{Prf}
The sparsity $A_{i,i}$ is $(i+1)sc$. The sparsity of $A_{0,1}$ is $sc(c+1)/2$, when $i\geq 1$, the sparsity of $A_{i,i+1}$ is $\max\{s(c-i),s(i+1)\}$. Therefore, the sparsity of matrix $A$ is $O(sc^2)$.
\end{Prf}

\begin{lemma}\label{lemma-4-2}
$\Vert A \Vert$ satisfies $\Vert A \Vert \leq (c+1)(\Vert F_1 \Vert+\Vert F_2 \Vert)$.
\end{lemma}
\begin{Prf}
By the definition of $A_{i,i}$, we have
\begin{equation}\label{norm-aii}
    \Vert A_{i,i}\Vert \leq (c+1)\Vert F_1\Vert,\quad i\in[c+1]_0,
\end{equation}
$[c+1]_0=[0,1,\dots,c]$ and in this paper, for any $i\in \mathbbm{N}$, $[i+1]_0=[0,1,\dots,i]$.
Then we analyze upper bound of $\Vert A_{i,i+1}\Vert $. We have $A_{0,1}A_{0,1}^T=\frac{c(c+1)}{2}F_2F_2^T$, then $\Vert A_{0,1} \Vert$ satisfies
\begin{eqnarray}\label{norm-a01}
    \Vert A_{0,1} \Vert= \sqrt{\frac{c(c+1)}{2}}\Vert F_2 \Vert < (c+1)\Vert F_2 \Vert.
\end{eqnarray} 
When $i\geq 1$, $A_{i,i+1}$ can be regarded as $\beta_i\times \beta_{i+1}$ dimensional block matrix, each block unit is an $n^{ i+1}\times n^{i+2}$ dimensional matrix and has the form $I^{\otimes j}\otimes F_2\otimes I^{\otimes i-j}$, $j\in [i+1]_0$.
From the structure of $A_{i,i+1}$, the number of non-zero block unit in each row or column of $A_{i,i+1}$ is no more than $c$, so $A_{i,i+1}$ can be divided into at most $c$ matrices which have only one non-zero block unit in each row or column. 
Therefore, 
\begin{equation}\label{norm-aii1}
    \Vert A_{i,i+1}\Vert \leq c\Vert F_2\Vert,\quad i\in[1,2,\dots,c-1].
\end{equation}
Combining Eq.(\ref{norm-aii}), Eq.(\ref{norm-a01}) and Eq.(\ref{norm-aii1}), $\Vert A \Vert$ satisfies
\begin{align}
    \Vert A \Vert\leq& \Vert {\rm diag}(A_{0,0},A_{1,1},\dots,A_{c,c}) \Vert+ \Vert {\rm diag}(A_{0,1},A_{1,2},\dots,A_{c-1,c}) \Vert \notag\\
    =&\max_{i=0}^{c}\{\Vert A_{i,i}\Vert \}+\max_{i=0}^{c-1}\{\Vert A_{i,i+1}\Vert \}\notag\\
    \leq& (c+1)(\Vert F_1 \Vert+\Vert F_2 \Vert).
\end{align}  
\end{Prf}

\begin{lemma}\label{lemma-4-3}
    The eigenvalue $\gamma_i$ of matrix $A$ satisfies $Re(\gamma_i)<0,i\in [N]_0$.
\end{lemma}
\begin{Prf}
From the structure of $A$, the eigenvalues of $A$ are the sets of eigenvalues of $A_{i,i}$ for $i\in[c+1]_0$. The eigenvalue of $A_{i,i}$ is the sum of $i+1$ eigenvalues of $F_1$. For any $j\in[1,2,\dots,n]$, eigenvalue of $F_1$ $\lambda_j$ satisfies $Re( \lambda_j)<0$, so the real part of all eigenvalues of $A_{i,i}$ is less than 0. Therefore, the eigenvalue $\gamma_i$ of $A$ satisfies $Re( \gamma_i)<0,i\in [N]_0$.
\end{Prf}

Next, we introduce the way to construct oracle $O_A$ of $A$. $O_A$ gives the way to extract non-zero element position and value of $A$.
We first give Lemma \ref{lemma-bm-new}.

\begin{lemma}\label{lemma-bm-new}
    Let matrix $B(m)=\sum_{j=0}^{m}{I_n^j\otimes F_1 \otimes I_n^{\otimes m-j}}$.
    Oracles $O_{m,1}$ and $O_{m,2}$ are defined as
    \begin{equation}
        \begin{aligned}
            &O_{m,1}|\vec{j}\rangle|l\rangle=|\vec{j}\rangle|g_m(\vec{j},l)\rangle,\\
            &O_{m,2}|\vec{j}\rangle|\vec{k}\rangle|z\rangle=|\vec{j}\rangle|\vec{k}\rangle|z\oplus B(m)_{\vec{j},\vec{k}}\rangle,\\
        \end{aligned}
    \end{equation}
    where $\vec{j}=j_{m}j_{m-1}\dots j_1j_0$, $j_i\in[n]_0$. $g_m(\vec{j},l)$ represents the column number of $l$-th non-zero element in $\vec{j}$-th row of $B(m)$, $g_m(\vec{j},l)$ is also written as $g_m(\vec{j},l)=k_mk_{m-1}\dots k_0$, $k_i\in[n]_0$.
    Then $O_{m,1}$, $O_{m,2}$ can be constructed by querying $O_{F1}$ $m+1$ times.
\end{lemma}
\begin{Prf}
When $m=0$, $B(0)=F_1$, $O_{0,1}$, $O_{0,2}$ can be constructed by querying $O_{F1}$ once.
Assuming when $m\geq 1$, Oracles $O_{m-1,1}$, $O_{m-1,2}$ of $B(m-1)$ are constructed by querying $O_{F1}$ $m$ times.
$B(m)$ is also written as
\begin{equation}
    B(m)=I\otimes B(m-1)+F_1\otimes I^{\otimes m}.
\end{equation}
Generally, we treat the diagonal element of $F_1$ as a non-zero element, so the sparsity of $B(m)$ is $(m+1)c-m$. $g_m(\vec{j},k)$ can be represented as
\begin{equation}\label{obm12}
    g_m(\vec{j},k)=\left\{\begin{array}{ll}
        f_1(j_m,k)j_{m-1}\dots j_0, & 0\leq k< g(j_m), \\
        j_mg_{m-1}(j_{m-1}\dots j_0,k-g(j_m)), & 0\leq k-g(j_m) \leq m(c-1)+1 ,\\
        f_1(j_m,k-m(c-1))j_{m-1}\dots j_0, & g(j_m)+1\leq k-m(c-1)\leq c-1,
        \end{array}\right.  
\end{equation}
where $f_1(j,k)$ and $g(j)$ are defined in Eq.(\ref{f1f2oracle}). Then $O_{m,1}$ can be constructed with Eq.(\ref{obm12}), the construction process needs to query $O_{m-1,1}$, $O_{F11}$, $O_{F13}$ once each.

On the other hand, the element of $B(m)$ is written as
\begin{equation}\label{bmjk}
    B(m)_{\vec{j},\vec{k}}=\delta_{j_mk_m}B(m-1)_{j_{m-1}\dots j_0,k_{m-1}\dots k_0}+\delta_{j_{m-1}\dots j_0,k_{m-1}\dots k_0}B_{j_m,k_m}.
\end{equation}
By Eq.(\ref{bmjk}), $O_{m,2}$ can be constructed by querying $O_{m-1,2}$ and $O_{F12}$ once each. Therefore $O_{m,1}$ and $O_{m,2}$ can be constructed by querying $O_{m-1,1}$, $O_{m-1,2}$, $O_{F11}$,$O_{F12}$ and $O_{F13}$ once each. 

From the above analysis, $O_{m,1}$ and $O_{m,2}$ can be constructed by querying $O_{F1}$ $m+1$ times.
\end{Prf}

\begin{lemma}\label{eq-oracle-oa}
The oracle $O_A$ of $A$ can be constructed by querying $O_{F1}$ $O(c)$ times and querying  $O_{F2}$ $O(1)$ times.
\end{lemma}
\begin{Prf}
To construct $O_A$, we need to construct oracles of $A_{i,i}$ and $A_{i,i+1}$.
We first consider $A_{i,i}$, we define $B(i)=\sum_{j=0}^{i}{I_n^j\otimes F_1 \otimes I_n^{\otimes c-i-j}}$,
by Lemma \ref{lemma-bm-new}, the oracle of $B(i)$ can be constructed by querying $O_{F1}$ $i+1$ times. By Eq.(\ref{eq-aii-def}), the oracle of $A_{i,i}$ can be constructed by querying oracle of $B(i)$ once.

Next we consider $A_{i,i+1}$. 
When $i=0$, $A_{0,1}=[F_2,F_2,\dots,F_2]$, the oracle of $A_{0,1}$ is constructed by querying $O_{F2}$ once. 
When $i\geq 1$, $A_{i,i+1}$ is also regarded as $\beta_i\times \beta_{i+1}$ dimensional block matrix $D(i)$, the dimension of each block unit is $n^{i+1}\times n^{i+2}$. 
The number of non-zero block unit in $j$-th row of $D(i)$ is $\sum_{k=0}^{i}{a_{i,j,k }}$. Consider the $l$-th non-zero bolck unit in $j$-th row of $D(i)$, 
$l$ is represented as $l=\sum_{k=0}^{i_1}{a_{i,j,k}}+i_2$, where $i_1\in [i]_0,i_2\in \mathbbm{N}$, the $l$-th non-zero block unit is $I^{\otimes i_1}\otimes F_2\otimes I^{\otimes i-i_1}$, and the corresponding $\vec{a}_{i+1,j^{'}}$ is represented as
\begin{equation}
    \vec{a}_{i+1,j^{'}}=[a_{i,j,0},\dots,a_{i,j,i_1},i_2,a_{i,j,i_1+1}-1-i_2,a_{i,j,i+2},\dots a_{i,j,i}],
\end{equation}
$j^{'}$ can be obtained from $\vec{a}_{i+1,j^{'}}$ with $O_{a1}$. The oracle of the non-zero block unit $I^{\otimes i_1}\otimes F_2\otimes I^{\otimes i-i_1}$ is constructed by querying $O_{F2}$ once. By realizing the above process with quantum circuit, we construct an oracle that extracts the non-zero position of $A_{i,i+1}$. The specific implementation process is
\begin{align}\label{Oiiplus1}
    &|j,b_i,b_{i-1},\dots ,b_{i_1},\dots ,b_0\rangle|l,l_b\rangle|0,0\rangle\notag\\
    \xrightarrow{O_{a2}}&|j,b_i,b_{i-1},\dots ,b_{i_1},\dots ,b_0\rangle|l,l_b\rangle|i_1,i_2\rangle\notag\\
    \xrightarrow{O_{a1}}&|j,b_i,b_{i-1},\dots ,b_{i_1},\dots ,b_0\rangle|j^{'},l_b\rangle|i_1,i_2\rangle\notag\\
    \xrightarrow{O_{F21}}&|j,b_i,b_{i-1},\dots ,b_{i_1},\dots ,b_0\rangle|j^{'},b_i,b_{i-1},\dots ,f_2(b_{i_1},l_b),\dots b_0\rangle|i_1,i_2\rangle\notag\\
    \xrightarrow{uncompute}&|j,b_i,b_{i-1},\dots ,b_{i_1},\dots ,b_0\rangle|j^{'},b_i,b_{i-1},\dots ,f_2(b_{i_1},l_b),\dots b_0\rangle|0,0\rangle.
\end{align}
There are some ancilla qubits to represent $|\vec{a}_{i,j}\rangle$ and $|\vec{a}_{i+1,j^{'}}\rangle$ in the process shown in Eq.(\ref{Oiiplus1}). For simplicity, we ignore the compute and uncompute process of $|\vec{a}_{i,j}\rangle$ and $|\vec{a}_{i+1,j^{'}}\rangle$.

The oracle that extracts the non-zero value of $A_{i,i+1}$ can also be constructed in a similar process. For any $j\in[\beta_i]_0$, $k\in[\beta_{i+1}]_0$, we can judge whether $D(i)_{j, k}$ is a non-zero block unit and use $l_{j,k}$ to represent the judgement. 
If $D(i)_{j, k}$ is a non-zero block unit, we can find $i_1$ and represent it as $I^{\otimes i_1}\otimes F_2\otimes I^{\otimes i-i_1}$, then we use $O_{F22}$ to extract the elements of the non-zero block unit. The whole process is shown as
\begin{align}\label{Oiiplus1-2}
    &|\vec{j}\rangle|\vec{k}\rangle|0,0\rangle|z\rangle=|j,b_i,b_{i-1},\dots ,b_0\rangle|k,b_i^{'},b_{i-1}^{'},\dots ,b_0^{'}\rangle|0,0\rangle|z\rangle\notag\\
    \xrightarrow{O_{a1},O_{a2}}&|\vec{j}\rangle|\vec{k}\rangle|l_{j,k},i_1\rangle|z\rangle\notag\\
    \xrightarrow{O_{F22}}&|\vec{j}\rangle|\vec{k}\rangle|l_{j,k},i_1\rangle|z\oplus l_{j,k}\times {F_2}_{b_{i_1},b_{i_1}^{'}} \prod_{l\in[i],l\neq i_1}{\delta_{b_lb_l^{'}}}\rangle\notag\\
    \xrightarrow{uncompute}&|\vec{j}\rangle|\vec{k}\rangle|0,0\rangle|z\oplus D(i)_{\vec{j},\vec{k}}\rangle.
\end{align}
The query complexity of $O_{F22}$ in the above process is $O(1)$. Therefore, oracle of $A_{i,i+1}$ can be constructed by querying $O_{F2}$ $O(1)$ times.

After constructing the oracles of $A_{i,i}$ and $A_{i,i+1}$, the oracle of $A$ can be directly constructed by querying oracles of $A_{i,i}$ and $A_{i,i+1}$ once. So the oracle $O_A$ can be constructed by querying $O_{F1}$ $O(c)$ times and querying  $O_{F2}$ $O(1)$ times.
\end{Prf}

\section{Condition Number}\label{sec-condition-num}

In this section, we give an upper bound of the condition number of $C_{m,k,p}(Ah)$ defined in Sect.\ref{sec-qldea}. We first analyze the upper bound of $\Vert e^{Aht} \Vert$, we have the following lemma.

\begin{lemma}\label{label-lem2}
    Consider the matrix $A$ defined in Sect.\ref{sec-combination}, when 
    \begin{equation}\label{eq-1027v2}
        \frac{(c+1)\Vert F_2 \Vert }{\vert Re(\lambda_1)\vert}\leq 1,
    \end{equation}
    for $t\geq 0$, $\Vert e^{At} \Vert$ satisfies
    \begin{equation}
        \Vert e^{At} \Vert\leq c+1.
    \end{equation}
\end{lemma}
\begin{Prf}
    We consider $A$ as a $c+1$-dimensional block matrix. $A$ is divided into 
    \begin{equation}\label{eq-1027v3}
        A=B+C,
    \end{equation}
    $B$ contains $A_{i,i}, i=0,1,\dots,c$ and  $C$ contains $A_{i,i+1}, i=0,1,\dots,c-1$. We analyze the upper bound of $\Vert e^{At}\Vert$ according to the method introduced in \cite{van1977sensitivity}, $e^{At}$ is written as
    \begin{equation}
        e^{(B+C)t}=e^{Bt}+\int_0^t{e^{B(t-t_0)}Ce^{(B+C)t_0}dt_0}.
    \end{equation}
    Using this formula to expand $e^{(B+C)t_0}$ we obtain 
    \begin{equation}
        e^{(B+C)t}=e^{Bt}+\int_0^t{e^{B(t-t_0)}Ce^{Bt_0}dt_0}+\int_0^t{\int_0^{t_0}{e^{B(t-t_0)}Ce^{B(t_0-t_1)}Ce^{(B+C)t_1}dt_1}dt_0}.
    \end{equation}
    Clearly, a repetition of this process gives 
    \begin{equation}
        e^{(B+C)t}=e^{Bt}+\sum_{k=0}^{c-1}{A_k(t)}+R_c(t),
    \end{equation}
    where 
    \begin{equation}
        A_{k}(t)=\int_{0}^{t} \int_{0}^{t_{0}} \cdots \int_{0}^{t_{k-1}} e^{B\left(t-t_{0}\right)} C e^{B\left(t_{0}-t_{1}\right)} C \cdots C e^{B t_{k}} d t_{k} \cdots d t_{0}
    \end{equation}
    and 
    \begin{equation}
        R_{c}(t)=\int_{0}^{t} \int_{0}^{t_{0}} \cdots \int_{0}^{t_{c-1}} e^{B\left(t-t_{0}\right)} C \cdots C e^{B\left(t_{c-1}-t_{c}\right)} C e^{(B+C) t_{c}} d t_{c} \cdots d t_{0}.
    \end{equation}
    Noting that 
    \begin{equation}
        \Vert A_k(t)\Vert \leq \Vert e^{Bt}\Vert \Vert C\Vert^{k+1}\times \frac{t^{k+1}}{(k+1)!}=\Vert e^{Bt}\Vert\frac{(\Vert Ct\Vert)^{k+1}}{(k+1)!}, k=0,1,\dots,c-1,
    \end{equation}
    \begin{equation}
        \Vert e^{Bt}\Vert=e^{Re(\lambda_1)t},\ \Vert C\Vert \leq (c+1)\Vert F_2\Vert,
    \end{equation}
    and the the matrix $[e^{B(t-t_0)}C]\dots [e^{B(t_{c-1}-t_c)}C]$ is zero beacuse it is the product of $c+1$, $(c+1)\times (c+1)$ strictly upper triangular block matrices and thus, $R_c(t)=0$. Hence, 
    \begin{equation}\label{eq-1027v1}
        \Vert e^{At} \Vert=\Vert e^{(B+C)t} \Vert \leq e^{Re(\lambda_1)t}\sum_{k=0}^{c}{\frac{((c+1)\Vert F_2\Vert t)^k}{k!}}.
    \end{equation}
    By Lemma \ref{lemma-new1}, Eq.(\ref{eq-1027v2}) and Eq.(\ref{eq-1027v1}),
    \begin{equation}
        \Vert e^{At} \Vert \leq c+1.
    \end{equation}
\end{Prf}

Then the upper bound of the condition number $\kappa_C$ of $C_{m,k,p}(Ah)$ is analyzed in Lemma \ref{theo-conditionnumber}.

\begin{lemma}\label{theo-conditionnumber}
    Consider the matrix $C_{m,k,p}(Ah)$ defined in Sect.\ref{sec-qldea}. Let $h\in \mathbbm{R^+}$ and satisfies $\Vert Ah \Vert \leq 1$, $c,m,k,p\in \mathbbm{Z}^{+}$. When $\frac{(c+1)\Vert F_2 \Vert }{\vert Re(\lambda_1)\vert}\leq 1$, $k\geq 5$ and $\frac{2}{(k+1)!}m(c+1)(c+2)\leq 1$, the condition number $\kappa_C$ of $C_{m,k,p}(Ah)$ satisfies 
    \begin{equation}
        \kappa_C\leq 2e\sqrt{k}(m(k+1)+p)(c+2),
    \end{equation}
    where $e$ is mathematical constant.
\end{lemma}

\begin{Prf}
    First we analyze the upper bound of $\Vert C_{m,k,p}(Ah)^{-1}\Vert$, we have
    \begin{equation}
        C_{m,k,p}(Ah)|x\rangle=|B\rangle,
    \end{equation}
    where $|B\rangle=\sum_{l=0}^{d}{b_l|l\rangle}$, $|l\rangle$ represents an $N$-dimensional state. We define $|b^l\rangle:=b_l|l\rangle$,
    \begin{equation}
        \sum_{l=0}^{d}{\Vert |b^l\rangle \Vert^2}=\Vert|B\rangle \Vert^2.
    \end{equation}
    For any $l\in[d]_0$, we define
    \begin{equation}
        |x^l\rangle:=C_{m,k,p}(Ah)^{-1}|b^l\rangle=\sum_{n=0}^{d}{|x_n^l\rangle}.
    \end{equation}

    We consider two cases: $0\leq l<m(k+1)$ and $m(k+1)\leq l\leq d$.

    When $0\leq l<m(k+1)$, assuming $l=a(k+1)+b$, $0\leq a<m$, $0\leq b\leq k$. Then based on definition of $x$, we have
    \begin{equation}\label{eq050705}
        \begin{aligned}
        |x_{i, j}^l\rangle &=0, & 0 \leq i<a, 0 \leq j \leq k, \\
        |x_{a, j}^l\rangle &=0, & 0 \leq j<b, \\
        |x_{a, j}^l\rangle &=b ! (Ah)^{j-b} / j!|b^l\rangle, & b \leq j \leq k \\
        |x_{a+1,0}^l\rangle &=T_{b, k}(Ah)|b^l\rangle, & \\
        |x_{a+1, j}^l\rangle &=(Ah)^{j} / j !|x_{a+1,0}^l\rangle, & \\
        |x_{a+2,0}^l\rangle &=T_{k}(Ah) |x_{a+1,0}^l\rangle=T_{k}(Ah) T_{b, k}(Ah)|b^l\rangle, & \\
        & \vdots & \\
        |x_{m, 0}^l\rangle &=T_{k}(Ah) |x_{m-1,0}^l\rangle=\left(T_{k}(Ah)\right)^{m-a-1} T_{b, k}(Ah)|b^l\rangle, & \\
        |x_{m, j}^l\rangle &=|x_{m, 0}\rangle=\left(T_{k}(Ah)\right)^{m-a-1} T_{b, k}(Ah)|b^l\rangle, & 1 \leq j \leq p,
        \end{aligned}
    \end{equation}
    where $|x_{i,j}^l\rangle=|x_{i(k+1)+j}^l\rangle$ and $T_{b,k}(Ah):=\sum_{j=b}^{k}{\frac{b!(Ah)^{j-b}}{j!}}$. By Lemma \ref{lemma-new7}, we have 
    \begin{equation}\label{eq050701}
        \Vert e^{Ah(m-a-1)}-T_k^{m-a-1}(Ah)\Vert \leq \frac{2}{(k+1)!}(m-a-1)(c+1)(c+2),
    \end{equation}
    by Lemma \ref{label-lem2}, $\Vert e^{Ah(m-a-1)}\Vert\leq c+1$, then we have
    \begin{equation}
        \Vert \left(T_{k}(Ah)\right)^{m-a-1}\Vert\leq c+2.
    \end{equation}
    Therefore,
    \begin{equation}\label{eq050704}
        \begin{aligned}
            \Vert (Ah)^{j-b}\Vert&\leq \Vert (Ah)\Vert^{j-b}\leq 1, \quad\quad\quad\quad\quad\quad\quad b \leq j \leq k, \\
            \Vert T_{b, k}(Ah) \Vert &\leq T_{b, k}(\Vert Ah\Vert) \leq e^{\Vert Ah\Vert} \leq e , \\
            \Vert T_{k}(Ah) \Vert &\leq T_{k}(\Vert Ah\Vert)\leq e^{\Vert Ah\Vert}\leq e ,  \\
            \Vert \left(T_{k}(Ah)\right)^{m-a-1} T_{b, k}(Ah) \Vert &\leq \Vert \left(T_{k}(Ah)\right)^{m-a-1}\Vert\Vert T_{b, k}(Ah)\Vert \leq e(c+2).
        \end{aligned}
    \end{equation}
    By Eq.(\ref{eq050705}) and Eq.(\ref{eq050704}), $\Vert x_{i,j}^l\Vert$ satisfies
    \begin{equation}
        \begin{aligned}
            \Vert|x_{a, j}^l\rangle \Vert &\leq\frac{b!}{j!}\Vert (Ah)^{j-b}\Vert \Vert|b^l\rangle \Vert\leq \Vert|b^l\rangle \Vert, & b \leq j \leq k \\
            \Vert|x_{a+1,0}^l\rangle\Vert &\leq\Vert T_{b, k}(Ah)\Vert \Vert|b^l\rangle\Vert\leq e\Vert|b^l\rangle \Vert , & \\
            \Vert|x_{i,j}^l\rangle\Vert &\leq e(c+2)\Vert|b^l\rangle \Vert , & a+1\leq i\leq m-1,0\leq j\leq k\\
            \Vert|x_{m,j}^l\rangle\Vert &\leq e(c+2)\Vert|b^l\rangle \Vert , & 0\leq j\leq p\\
        \end{aligned}
    \end{equation}
    therefore,
    \begin{equation}\label{eq-condition-1}
        \Vert C_{m,k,p}(Ah)^{-1}|b^l\rangle\Vert^2\leq (m(k+1)+p)(e(c+2)))^2\Vert|b^l\rangle \Vert^2.
    \end{equation}

    When $m(k+1)\leq l \leq d$, assuming $l=m(k+1)+b$, $0\leq b\leq p$, $x_{i,j}$ satisfies 
    \begin{equation}
        \begin{aligned}
        |x_{i, j}^l\rangle &=0, & 0 \leq i<m, 0 \leq j \leq k, \\
        |x_{m, j}^l\rangle &=0, & 0 \leq j<b, \\
        |x_{m, j}^l\rangle &=|b^l\rangle, & b \leq j \leq p, 
        \end{aligned}
    \end{equation}
    then 
    \begin{equation}\label{eq-condition-2}
        \Vert C_{m,k,p}(Ah)^{-1}|b^l\rangle\Vert^2\leq p\Vert|b^l\rangle \Vert^2.
    \end{equation}

    From Eq.(\ref{eq-condition-1}) and Eq.(\ref{eq-condition-2}), for any $|B\rangle$, $\Vert C_{m,k,p}(Ah)^{-1}|B\rangle\Vert$ satisfies
    \begin{align}
        \Vert C_{m,k,p}(Ah)^{-1}|B\rangle\Vert^2\leq& (m(k+1)+p)\sum_{l=0}^{d}{\Vert C_{m,k,p}(Ah)^{-1}|b^l\rangle\Vert^2}\notag\\
        \leq& (m(k+1)+p)^2(e(c+2))^2\Vert |B\rangle\Vert^2,
    \end{align}
    then 
    \begin{equation}\label{C_A_inverse}
        \Vert C_{m,k,p}(Ah)^{-1}\Vert \leq (m(k+1)+p)e(c+2).
    \end{equation}
    From Lemma 4 in \cite{berry2017quantum}, we have
    \begin{equation}\label{C_A_norm}
        \Vert C_{m,k,p}(Ah)\Vert\leq 2\sqrt{k}.
    \end{equation}
    Therefore, by Eq.(\ref{C_A_norm}) and Eq.(\ref{C_A_inverse}), we have $\kappa_C\leq 2e\sqrt{k}(m(k+1)+p)(c+2)$.
\end{Prf}

\section{Solution Error}\label{sec-solution-error}

In this section, we analyze the solution error of our algorithm. The error mainly comes from two aspects: (1) Homotopy perturbation method truncation error. The solution $\tilde{u}(t)$ defined in Eq.(\ref{exp_solution}) is an approximate solution of Eq.(\ref{nonlinear_eq}), the error bound is determined by the truncation order $c$, in Sect.\ref{subsec-hpm-error}, we analyze the convergence condition of Eq.(\ref{exp_solution}) and give the error bound. (2) Linear ODEs solution error. We solve the linear ODEs defined in Eq.(\ref{eq-linear}) with the quantum algorithm proposed in \cite{berry2017quantum}. This algorithm also generates intermediate error, we analyze the error bound in Sect.\ref{subsec-ode-error}.

\subsection{Homotopy perturbation method truncation error}\label{subsec-hpm-error}
We first analyze homotopy perturbation method truncation error, the result is shown in Lemma \ref{theo-error-1}.
\begin{lemma}\label{theo-error-1}
    Consider the nonlinear ODEs defined in Eq.(\ref{nonlinear_eq}), the solution of Eq.(\ref{nonlinear_eq}) obtained by homotopy perturbation method is written as $\tilde{u}(t)=\sum_{i=0}^{c}{\nu_i(t)}$, $u(t)$ represents the exact solution of Eq.(\ref{nonlinear_eq}). When $K<1$ and $c>\log_{1/K}{\frac{1}{\epsilon(1-K)}}$, $\tilde{u}(t)$ satisfies
    \begin{equation}\label{error-1-con}
        \Vert u(t)-\tilde{u}(t)\Vert \leq \epsilon.
    \end{equation}
\end{lemma}
\begin{Prf}
    $u(t)$ can be represented as $u(t)=\sum_{i=0}^{\infty}{\nu_i(t)}$, Eq.(\ref{error-1-con}) is transformed into 
    \begin{equation}\label{u_error}
        \Vert\sum_{j=c+1}^{\infty}{\nu_j(t)}\Vert<\epsilon.
    \end{equation}
    To prove Eq.(\ref{u_error}), we analyze the upper bound of $\Vert \nu_i\Vert$ defined in Eq.(\ref{linear_differential}). $\nu_0(t)=e^{F_1t}u_{in}$, we have
    \begin{equation}
        \Vert\nu_0(t)\Vert\leq\Vert e^{F_1t}\Vert\times\Vert u_{in}\Vert\leq \Vert u_{in}\Vert.
    \end{equation}
    We define $K_1=\frac{\Vert u_{in} \Vert \Vert F_2 \Vert}{\vert Re(\lambda_1) \vert}$ and
    assume when $j\leq i$, $\Vert \nu_j \Vert$ satisfies
    \begin{equation}\label{error-050801}
        \Vert\nu_j\Vert\leq \alpha_jK_1^j\Vert u_{in} \Vert,\ j=0,1,\dots ,i.
    \end{equation}
    Then $\Vert\nu_{i+1}(t)\Vert$ satisfies
    \begin{align}\label{error-3}
        \Vert\nu_{i+1}(t)\Vert&=\Vert\int_{0}^{t}{e^{F_1(t-\tau)}F_2\sum_{j=0}^{i}{\nu_j(\tau)\otimes\nu_{i-j}(\tau)}d\tau}\Vert\notag\\
        &\leq\int_{0}^{t}{\Vert e^{F_1(t-\tau)}F_2\Vert\times (\sum_{j=0}^{i}{\alpha_j\alpha_{i-j}})K_1^{i}\Vert u_{in}\Vert^2d\tau}\notag\\
        &\leq (\sum_{j=0}^{i}{\alpha_j\alpha_{i-j}})K_1^{i}\Vert u_{in}\Vert^2\Vert F_2\Vert\int_{0}^{t}{\Vert e^{Re(\lambda_1)(t-\tau)}\Vert d\tau}\notag\\
        &\leq (\sum_{j=0}^{i}{\alpha_j\alpha_{i-j}})K_1^{i}\Vert u_{in}\Vert^2\Vert F_2\Vert\frac{1-e^{Re(\lambda_1)t}}{|Re(\lambda_1)|}\notag\\
        &\leq (\sum_{j=0}^{i}{\alpha_j\alpha_{i-j}})K_1^{i+1}\Vert u_{in} \Vert.
    \end{align}
    By Eq.(\ref{error-050801}), Eq.(\ref{error-3}), $\alpha_i$ can be defined as
    \begin{equation}\label{catalan-seq}
        \alpha_{i+1}=\sum_{j=0}^{i}{\alpha_j\alpha_{i-j}},\ \alpha_0=1.
    \end{equation}
    Eq.(\ref{catalan-seq}) is the catalan sequence\cite{koshy2008catalan} and satisfies
    \begin{equation}\label{eq-catalan}
        \alpha_i=\frac{1}{i+1} \binom{2i}{i}\approx\frac{4^i}{i^{3/2}\sqrt{\pi}}<4^i.
    \end{equation}
    Combining Eq.(\ref{parameter-r-new}), Eq.(\ref{error-050801}), Eq.(\ref{error-3}), Eq.(\ref{catalan-seq}) and Eq.(\ref{eq-catalan}), for any $i\in \mathbbm{N}$, $\Vert\nu_{i}(t)\Vert$ has the upper bound
    \begin{equation}\label{nu_i_norm}
        \Vert\nu_{i}(t)\Vert< (4K_1)^i \Vert u_{in} \Vert\leq K^{i+1}.
    \end{equation}
    Substituting Eq.(\ref{nu_i_norm}) into Eq.(\ref{u_error}), we have
    \begin{equation}\label{error-050802}
        \sum_{i=c+1}^{\infty}{K^{i+1}}<\epsilon.
    \end{equation}
    Therefore, when $K<1$ and $c>\log_{1/K}{\frac{1}{\epsilon(1-K)}}$, we have $\Vert u(t)-\tilde{u}(t)\Vert \leq \epsilon$.
\end{Prf}

\subsection{Linear ODEs solution error}\label{subsec-ode-error}

Then we analyze the error of solving the linear ODEs defined in Eq.(\ref{eq-linear}), we have a similar conclusion with Theorem $\bm{6}$ in \cite{berry2017quantum}, the difference comes from the upper bound of $\Vert e^{Aj}\Vert$ or $\Vert T_k^j(A)\Vert$ in our work is different from their work. Our result is shown in Lemma \ref{theo-1}.

\begin{lemma}\label{theo-1}
    Consider the linear ODEs defined in Eq.(\ref{eq-linear}) and the system of linear equations defined in Eq.(\ref{3-30-2}). Let $h\in \mathbbm{R}^+$ satisfies $\Vert Ah \Vert\leq 1$. When $\frac{2}{(k+1)!}m(c+1)(c+2)\leq 1$, we have
    \begin{equation}
        \Vert |y(jh)\rangle-|x_{j,0}\rangle\Vert\leq \frac{2j(c+1)(c+2)\Vert y_{in} \Vert}{(k+1)!}, j=0,1,\dots,m.
    \end{equation}
\end{lemma}
\begin{Prf}
    The exact solution of Eq.(\ref{eq-linear}) at $t=jh$ is $|y(jh)\rangle$, it is written as
    \begin{equation}
        |y(jh)\rangle=e^{Ajh}|y(0)\rangle.
    \end{equation}
    The solution of Eq.(\ref{3-30-2}) $|x_{j,0}\rangle$ satisfies
    \begin{equation}
        |x_{j,0}\rangle=T_k^{j}(Ah)|x_{0,0}\rangle,
    \end{equation}
    where $T_k(Ah)=\sum_{l=0}^{k}{\frac{(Ah)^l}{l!}}$, the initial value satisfies $|y(0)\rangle=|x_{0,0}\rangle=|y_{in}\rangle$. By Lemma \ref{label-lem2} and Lemma \ref{lemma-new7}, $\Vert |y(jh)\rangle-|x_{j,0}\rangle\Vert$ satisfies
    \begin{align}
        \Vert |y(jh)\rangle-|x_{j,0}\rangle\Vert\leq \Vert e^{Ajh}-T_k^{j}(Ah) \Vert \Vert y_{in}\Vert \leq \frac{2j(c+1)(c+2)\Vert y_{in} \Vert}{(k+1)!},\ j=0,1,\dots,m.
    \end{align}
\end{Prf}

\section{Success Probability}\label{sec-success-prob}

As introduced in Sect.\ref{sec-measurement}, there are two probabilistic steps in our method. This section gives a lower bound of the success probability of these two steps. The results are shown in Lemma \ref{theo-success-1} and Lemma \ref{theo-success-2}.

Firstly, we analyze the success probability of getting $|x_{m,i}\rangle,i=0,1,\dots ,p$ when measuring $|x\rangle$. By setting appropriate conditions, Lemma \ref{theo-success-1} gives the same conclusion as Theorem $\bm{7}$ in \cite{berry2017quantum}.

\begin{lemma}\label{theo-success-1}
    Consider the system of linear equations defined in Eq.(\ref{3-30-2}). Let $h\in \mathbbm{R}^+$ satisfies $\Vert Ah \Vert\leq 1$. $g=\max_{t\in [0,mh]}\Vert |y(t)\rangle\Vert /\Vert |y(mh)\rangle \Vert$, $m,k,p\in \mathbbm{Z}^{+}$. When $(k+1)!\geq 50m(c+1)(c+2)g$, we have
    \begin{equation}
        \frac{\Vert |x_{m,0}\rangle\Vert^2}{\Vert |x\rangle\Vert^2}\geq \frac{1}{p+77mg^2}.
    \end{equation}
\end{lemma}
\begin{Prf}
    As introduced before, $|x_{m,j}\rangle=|x_{m,0}\rangle$ for $j\in[p]_0$, we define
    \begin{equation}
        |x_{good}\rangle:=\sum_{j=0}^{p}|m(k+1)+j\rangle|x_{m,j}\rangle
    \end{equation}
    and
    \begin{equation}
        |x_{bad}\rangle:=\sum_{i=0}^{m-1}{\sum_{j=0}^k{|i(k+1)+j\rangle|x_{i,j}\rangle}}.
    \end{equation}
    We see $|x\rangle=|x_{good}\rangle+|x_{bad}\rangle$ and $\langle x_{good}|x_{bad}\rangle=0$, then 
    \begin{align}
        \Vert |x\rangle\Vert^2=&\Vert |x_{good}\rangle\Vert^2+\Vert |x_{bad}\rangle\Vert^2\notag\\
        =&(p+1)\Vert |x_{m,0}\rangle\Vert^2+\Vert |x_{bad}\rangle\Vert^2.
    \end{align}
    Next we give a lower bound of $\Vert |x_{m,0}\rangle\Vert$ and an upper bound of $\Vert |x_{bad}\rangle\Vert$. We define $q=\Vert |y(mh)\rangle\Vert$,
    by Lemma \ref{theo-1} and $(k+1)!\geq 50m(c+1)(c+2)g$, we have
    \begin{equation}
        \Vert |x_{i,0}\rangle-|y(ih)\rangle\Vert \leq 0.04q,\  0\leq l\leq m.
    \end{equation}
    By the definition of $g$, $\Vert |y(ih)\rangle\Vert\leq gq$ for any $i\in[m-1]_0$, then we have
    \begin{equation}
        \Vert |x_{i,0}\rangle\Vert \leq (g+0.04)q\leq 1.04gq,\  0\leq i \leq m-1,
    \end{equation}
    and
    \begin{equation}\label{eq-succ-050801}
        0.96q\leq \Vert |x_{m,0}\rangle\Vert \leq 1.04q.
    \end{equation}
    For any $i\in[m]_0$, $|x_{i,j}\rangle=(Ah/j)|x_{i,j-1}\rangle$, we have
    \begin{equation}
        |x_{i,j}\rangle=\frac{(Ah)^{j-1}}{j!}|x_{i,1}\rangle, \ 2\leq j\leq k.
    \end{equation}
    We have $\Vert Ah\Vert\leq 1$, therefore, 
    \begin{equation}
        \Vert |x_{i,j}\rangle\Vert \leq \frac{\Vert |x_{i,1}\rangle\Vert}{j!},\  2\leq j\leq k.
    \end{equation}
    Next, based on $|x_{i+1,0}\rangle=|x_{i,0}\rangle+\sum_{j=1}^{k}{|x_{i,j}\rangle}$, we have
    \begin{align}
        2.08gq\geq & \Vert |x_{i+1,0}\rangle\Vert+\Vert |x_{i,0}\rangle\Vert\notag\\
        \geq &\Vert |x_{i+1,0}\rangle-|x_{i,0}\rangle\Vert\notag\\
        \geq &\Vert |x_{i,1}\rangle\Vert-\sum_{j=2}^{k}{\Vert |x_{i,j}\rangle\Vert}\notag\\
        \geq & \left(1-\sum_{j=2}^k{\frac{1}{j!}}\right)\Vert |x_{i,1}\rangle\Vert\notag\\
        \geq & (3-e)\Vert |x_{i,1}\rangle\Vert,
    \end{align}
    then
    \begin{equation}
        \Vert |x_{i,1}\rangle\Vert\leq \frac{2.08gq}{3-e},\ 0\leq i\leq m-1.
    \end{equation}
    and
    \begin{equation}
        \Vert |x_{i,j}\rangle\Vert\leq \frac{2.08gq}{j!(3-e)},\ 0\leq i\leq m-1,\ 1\leq j\leq k.
    \end{equation}
    $\Vert |x_{bad}\rangle\Vert$ satisfies
    \begin{align}\label{eq-success-4}
        \Vert |x_{bad}\rangle\Vert^2 & =\sum_{i=0}^{m-1}{\Vert |x_{i,0}\rangle\Vert^2}+\sum_{i=0}^{m-1}{\sum_{j=1}^k{\Vert |x_{i,j}\rangle\Vert^2}}\notag\\
        &\leq 1.04^2mg^2q^2+m\sum_{j=1}^k{\frac{(2.08gq)^2}{(j!)^2(3-e)^2}}\notag\\
        &\leq 70.9mg^2q^2.
    \end{align}
    The last step of Eq.(\ref{eq-success-4}) is derived from the inequality $\sum_{j=1}^k{\frac{1}{(j!)^2}}\leq I_0(2)-1<1.28$,
    where $I_0(2)<2.28$ a modified Bessel function of the first kind\cite{berry2017quantum}.
    Combining Eq.(\ref{eq-succ-050801}) and Eq.(\ref{eq-success-4}), we have 
    \begin{align}
        \frac{\Vert |x_{m,0}\rangle\Vert^2}{\Vert |x\rangle\Vert^2}&\geq \frac{(0.96q)^2}{p(0.96q)^2+70.9mg^2q^2}\notag\\
        &\geq \frac{1}{p+77mg^2}.
    \end{align}
\end{Prf}

Secondly, we analyze the success probability of the second probabilistic step. After the first measurement, the desired state is the state $|y(T)\rangle$ defined in Eq.(\ref{def-vec-x}). Then we measure the first qubit register of $|y(T)\rangle$, if the result is $|0,0\rangle$, we have a state $\epsilon$-close to $|u(T)/\Vert u(T)\Vert\rangle$ in the second qubit register of $|y(T)\rangle$. The lower bound of the success probability in this measurement is analyzed in Lemma \ref{theo-success-2}.
\begin{lemma}\label{theo-success-2}
    Let $T\in \mathbbm{R}^{+}$, $\eta^{'}=K/\Vert \tilde{u}(T)\Vert$. When $K< \sqrt{2}/2$, we have
    \begin{equation}
        \frac{\Vert |y_0(T)\rangle\Vert^2}{\Vert |y(T)\rangle\Vert^2}\geq\frac{1-2K^2}{1-2K^2+2(\eta^{'})^2}.
    \end{equation}
\end{lemma}
\begin{Prf}
    We rearrange the components of $\vec{y}$,
    \begin{equation}
        \vec{y}=[\vec{y}_0^{'},\vec{y}_1^{'},\dots ,\vec{y}_c^{'}],
    \end{equation}
    where $\vec{y}_0^{'}=\vec{y}_0=\sum_{j=0}^c{\nu_j}$, when $i\geq 1$, the component of $\vec{y}_i^{'}$ is represented as $\nu_{a_{i,0}^{'}}\otimes \nu_{a_{i,1}^{'}}\otimes \dots \otimes \nu_{a_{i,k}^{'}}$, $a_{i,j}^{'}$ satisfies
    \begin{equation}\label{eq-success-2-1}
        k\geq 1,\ a_{i,j}^{'}\geq 0,\ \sum_{j=0}^{k}{(a_{i,j}^{'}+1)}=i+1.
    \end{equation}
    By Eq.(\ref{eq-success-2-1}), the number of elements in $\vec{y}_i^{'}$ is $2^i-1$. By Eq.(\ref{nu_i_norm}) and Eq.(\ref{eq-success-2-1}), 
    for any $j\in [2^i-1]_0$, we have $\Vert \vec{y}_{i,j}^{'} \Vert \leq K^{i+1}$. Therefore,
    \begin{equation}\label{eq060201}
        \Vert \vec{y}_i^{'}\Vert^2 \leq (2^i-1)K^{2(i+1)}< (2K^2)^i.
    \end{equation}
    By Eq.(\ref{eq060201}) and $\Vert \vec{y}_0\Vert=\Vert \vec{y}_0^{'}\Vert=\Vert \tilde{u}(T) \Vert = K/\eta^{'}$, we have
    \begin{align}
        \frac{\Vert |y_0\rangle\Vert^2}{\Vert |y\rangle\Vert^2}=& \frac{(K/\eta^{'})^2}{(K/\eta^{'})^2+\sum_{i=1}^{c}{ (2K^2)^i} }\\
        \geq &\frac{(K/\eta^{'})^2}{(K/\eta^{'})^2+\frac{2K^2}{1-2K^2} }\\
        = &\frac{1-2K^2}{1-2K^2+2(\eta^{'})^2}.
    \end{align}
\end{Prf}

\section{Main Result}\label{sec-main-result}

In this section, we give the main result of our work.

\begin{Theo}\label{main_result}

    Given $n$-dimensional nonlinear dissipative ODEs $\frac{du}{dt}=F_1u+F_2u^{\otimes2}$ defined in Eq.(\ref{nonlinear_eq}) and construct the linear ODEs $\frac{d\vec{y}}{dt}=A\vec{y}$ defined in Eq.(\ref{eq-linear}). Let $T>0$, $g=\max_{t\in [0,T]}\Vert |y(t)\rangle\Vert /\Vert |y(T)\rangle \Vert$, $\eta=\Vert u_{in}\Vert/\Vert u(T)\Vert$, $c=\left\lceil \log_{1/K}{\frac{4\Vert u_{in}\Vert }{(1-K)\epsilon \eta}}\right\rceil$. 
    When $K< \sqrt{2}/2$ and $\frac{(c+1)\Vert F_2 \Vert }{\vert Re(\lambda_1)\vert}\leq 1$,
    there exists a quantum algorithm to produce $|u_{out}(T)\rangle$ which satisfies $\Vert u_{out}(T)-u(T)/\Vert u(T)\Vert\Vert \leq \epsilon$ with $\Omega(1)$ success probability. The query complexity of the algorithm for $O_{F1}$, $O_{F2}$, $O_{u}$ is 
    \begin{equation}
        O\left(\frac{g\eta s T(\Vert F_1\Vert+\Vert F_2 \Vert)}{\sqrt{1-2K^2}\Vert u_{in}\Vert}{\rm poly}\left(\log\left(\frac{g\eta sT\Vert F_1\Vert\Vert F_2 \Vert}{\epsilon(1-2K^2)\Vert u_{in}\Vert}\right)\right)\right).
    \end{equation}
    The gate complexity of this algorithm is larger than its query complexity by a factor of 
    \begin{equation}
        O\left({\rm poly}\left(\log\left(\frac{ng\eta sT\Vert F_1\Vert\Vert F_2 \Vert}{\epsilon(1-2K^2)\Vert u_{in}\Vert}\right)\right)\right).
    \end{equation}
\end{Theo}

\begin{Prf}

    \textbf{Whole process.} First we show the whole process of our algorithm.
    We define $\eta^{'}=\eta K/\Vert u_{in}\Vert$ and set 
    \begin{equation}
        \epsilon\leq 0.1\sqrt{1-2K^2}/\eta^{'},\ \epsilon_1=\frac{\epsilon K}{4\eta^{'}},\ \delta=\epsilon\frac{\sqrt{1-2K^2}}{30\sqrt{78m}g\eta^{'}},
    \end{equation}
    and construct the $N$-dimensional linear ODEs defined in Eq.(\ref{eq-linear})
    \begin{equation}
        d\vec{y}/dt=A\vec{y},\ \vec{y}(0)=y_{in}.
    \end{equation}
    We also set $h=T/\lceil T\Vert A\Vert \rceil$, $m=p=T/h=\lceil T\Vert A\Vert\rceil$, $k=\lfloor\frac{2\log(\Omega)}{\log(\log(\Omega))}\rfloor$,
    where $\Omega=50m(c+1)(c+2)g/\delta$, so $k$ satisfies $(k+1)!\geq \Omega$. 
    Then we construct the linear system defined in Eq.(\ref{3-30-2}) and solve the linear system with the algorithm proposed in \cite{childs2017quantum}. The normalized solution of Eq.(\ref{3-30-2}) is represented as
    \begin{equation}
        |x\rangle=\sum_{l=0}^{d}{|l\rangle|x_l\rangle},
    \end{equation}
    where $d=m(k+1)+p$. $|x\rangle$ can also be represented as 
    \begin{equation}\label{eq-main-051202}
        |\bar{x}\rangle=\sum_{l=0}^{d}{\alpha_l|l\rangle|\bar{x}_l\rangle},
    \end{equation}
    where $|\bar{x}_l\rangle$ is a normalized state. Assuming the output state of quantum linear system algorithm\cite{childs2017quantum} is  
    \begin{equation}
        |\bar{x}^{'}\rangle=\sum_{l=0}^{d}{\alpha_l^{'}|l\rangle|\bar{x}_l^{'}\rangle}.
    \end{equation}
    By Theorem $\bm 5$ in \cite{childs2017quantum}, we make $|\bar{x}^{'}\rangle$ satisfies
    \begin{equation}\label{eq-main-051204}
        \Vert |\bar{x}\rangle-|\bar{x}^{'}\rangle \Vert \leq \delta.
    \end{equation}
    Then we execute measurement step discussed in Sect.\ref{sec-measurement} and get a state $\epsilon$-close to $|u(T)/\Vert u(T)\Vert\rangle$ with success probability $O((1-2K^2)/(g\eta^{'})^2)$. We can amplify the success probability to $\Omega(1)$ with quantum amplitude amplification algorithm\cite{brassard2002quantum} by running quantum linear system algorithm $O(g\eta^{'}/\sqrt{1-2K^2})$ times.

    \textbf{Proof of correctness.} Then we analyze the error bound and the impact of error on success probability.

    Assuming $u(T)$ represents the exact solution. We define $|\tilde{u}(T)\rangle$ as
    \begin{equation}
        |\tilde{u}(T)\rangle=|\sum_{i=0}^{c}{\nu_i(T)}\rangle=|y_{0,0}(T)\rangle.
    \end{equation}
    By Lemma \ref{theo-error-1} and our choice of parameter $c$, we have  
    \begin{equation}
        \Vert |u(T)\rangle-|\tilde{u}(T)\rangle\Vert \leq \epsilon_1.
    \end{equation}
    Then by Lemma \ref{berry-lemma13} and $\Vert |u(T)\rangle\Vert = K/\eta^{'}$, we have 
    \begin{equation}\label{main-0510-05}
        \Vert |\bar{u}(T)\rangle-|\bar{\vec{y}}_{0,0}(T)\rangle \Vert \leq \frac{2\eta^{'}\epsilon_1}{K}=\epsilon/2.
    \end{equation}
    Let $S:=\{m(k+1),m(k+1)+1,\dots ,m(k+1)+p\}$. By Lemma \ref{theo-1} and our choice of parameter $k$, for any $l\in S$, we have
    \begin{equation}\label{eq-main-051201}
        \Vert |x_l\rangle-|y(T)\rangle\Vert \leq \delta \Vert |y(T)\rangle\Vert.
    \end{equation}
    By Eq.(\ref{eq-main-051201}) and Lemma \ref{berry-lemma13}, we have
    \begin{equation}\label{eq060203}
        \Vert |\bar{x}_l\rangle-|\bar{y}(T)\rangle\Vert \leq 2\delta.
    \end{equation}
    By Lemma \ref{theo-success-1}, for any $l\in S$, we have
    \begin{equation}\label{eq-main-051203}
        \alpha_l=\frac{\Vert |x_l\rangle\Vert}{\Vert |x\rangle \Vert}\geq \frac{1}{\sqrt{78m}g}\geq \delta.
    \end{equation}
    By Eq.(\ref{eq-main-051204}), Eq.(\ref{eq-main-051203}) and Lemma \ref{berry-lemma14}, we have 
    \begin{equation}\label{eq060202}
        \Vert |\bar{x}_l\rangle-|\bar{x}^{'}_l\rangle \Vert \leq \frac{2\delta}{\alpha_l-\delta}.
    \end{equation}
    Then, combine Eq.(\ref{eq060203}) and Eq.(\ref{eq060202}), we have
    \begin{align}\label{main-0510-02}
        \Vert |\bar{x}^{'}_l\rangle-|\bar{y}(T)\rangle \Vert&\leq \Vert |\bar{x}_l^{'}\rangle-|\bar{x}_l\rangle\Vert+\Vert |\bar{x}_l\rangle-|\bar{y}(T)\rangle\Vert\notag\\
        &\leq 2\delta\left(1+\frac{1}{\alpha_l-\delta}\right).
    \end{align}
    We default $\alpha_l$ is small enough, such as $\alpha_l<0.5$, then by Eq.(\ref{eq-main-051203}) and $\delta=\epsilon\frac{\sqrt{1-2K^2}}{30\sqrt{78m}g\eta^{'}}$, we have 
    \begin{equation}\label{eq115}
        2\delta(1+\frac{1}{\alpha_l-\delta})\leq 2\delta\frac{\sqrt{3}}{\alpha_l}\leq \frac{\epsilon \sqrt{1-2K^2}}{5\sqrt{3}\eta^{'}}.
    \end{equation}
    $|y(T)\rangle$ and $|\bar{x}_l^{'}\rangle$ are also written as 
    \begin{equation}
        |\bar{y}(T)\rangle=\sum_{w=0}^{c}{\chi_w|\bar{y}_w(T)\rangle}
    \end{equation}
    and 
    \begin{equation}
        |\bar{x}_l^{'}\rangle=\sum_{w=0}^{c}{\chi_w^{'}|\bar{x}_{l,w}^{'}\rangle}.
    \end{equation}
    By Lemma \ref{theo-success-2}, we have 
    \begin{equation}\label{main-0510-01}
        \chi_0=\frac{\Vert |y_0(T)\rangle \Vert}{\Vert |y(T)\rangle \Vert }\geq \sqrt{\frac{1-2K^2}{(1-2K^2)+2(\eta^{'})^2}}\geq \frac{\sqrt{1-2K^2}}{\sqrt{3}\eta^{'}}.
    \end{equation}
    By Eq.(\ref{main-0510-02}), Eq.(\ref{eq115}), Eq.(\ref{main-0510-01}) and Lemma \ref{berry-lemma14}, we have 
    \begin{equation}\label{main-0510-03}
        \Vert |\bar{x}_{l,0}^{'}\rangle- |\bar{y}_{0,0}(T)\rangle \Vert \leq \frac{2(\Vert |\bar{x}^{'}_l\rangle-|\bar{y}(T)\rangle \Vert)}{\chi_0-(\Vert |\bar{x}^{'}_l\rangle-|\bar{y}(T)\rangle \Vert)}\leq \frac{2\epsilon}{5-\epsilon}\leq \epsilon/2.
    \end{equation}
    We notice $|\bar{x}_{l,0}^{'}\rangle$ is the output state of our algorithm, we have $|u_{out}(T)\rangle=|\bar{x}_{l,0}^{'}\rangle$.
    Combining Eq.(\ref{main-0510-05}) and Eq.(\ref{main-0510-03}), we have
    \begin{equation}
        \Vert |\bar{u}(T)\rangle-|u_{out}(T)\rangle  \Vert\leq  \Vert |\bar{u}(T)\rangle-|\bar{y}_{0,0}(T)\rangle \Vert+\Vert |\bar{x}_{l,0}^{'}\rangle- |\bar{y}_{0,0}(T)\rangle \Vert\leq \epsilon.
    \end{equation}

    On the other hand, caused by solution error, the success probability also changes.
    By Lemma \ref{berry-lemma15}, Eq.(\ref{main-0510-01}) and $\epsilon\leq 0.1\sqrt{1-2K^2}/\eta^{'}$, we have 
    \begin{equation}
        \chi_0^{'}\geq \chi_0-\frac{1}{\sqrt{3}\eta^{'}}\sqrt{1-2K^2}\times \frac{1}{5}\epsilon \geq \chi_0-\epsilon/2\geq \frac{\sqrt{1-2K^2}}{2\eta^{'}}
    \end{equation}
    and
    \begin{equation}
        \alpha_l^{'}\geq \alpha_l-\delta \geq \left(\frac{1}{\sqrt{78}}-\frac{1}{30\sqrt{78}}\right)\frac{1}{\sqrt{m}g}\geq\frac{1}{10\sqrt{m}g},
    \end{equation}
    then 
    \begin{equation}
        \sum_{l\in S}{\vert \alpha_l^{'}\vert^2}\geq \frac{p}{100mg^2}=\frac{1}{100g^2},
    \end{equation}
    Therefore, the success probability of our algorithm is 
    \begin{equation}
        p=(\chi_0^{'})^2\times \left(\sum_{l\in S}{\vert \alpha_l^{'}\vert^2}\right)\geq \frac{1-2K^2}{400(g\eta^{'})^2}.
    \end{equation}

    \textbf{Complexity analysis.}
    Finally, we analyze the complexity of our algorithm. 

    By Lemma \ref{lemma-4-1} and Lemma \ref{lemma-4-2}, the sparsity of $A$ is $c^2s$ and $\Vert A\Vert \leq (c+1)(\Vert F_1 \Vert+\Vert F_2 \Vert)$. The sparsity $s_C$ of $C_{m,k,p}(Ah)$ satisfies
    \begin{equation}\label{eq-sc}
        s_C<k+c^2s.
    \end{equation}
    By Lemma \ref{theo-conditionnumber}, the condition number $\kappa_C$ of $C_{m,k,p}(Ah)$ satisfies
    \begin{equation}\label{eq-kappac}
        \kappa_C\leq 2e\sqrt{k}((m(k+1)+p))(c+2).
    \end{equation}
    By Theorem 5 in \cite{childs2017quantum} and Eq.(\ref{eq-main-051204}), the query complexity of quantum linear system algorithm to oracle of $C_{m,k,p}(Ah)$ and $|0\rangle|y_{in}\rangle$ is 
    \begin{equation}\label{eq-qlsa-complexity}
        O(s_C\kappa_C{\rm poly}({\rm log}(s_C\kappa_C/\delta))).
    \end{equation}
    
    By the definition of $C_{m,k,p}(Ah)$, the oracle of $C_{m,k,p}(Ah)$ can be constructed by querying oracle $O_A$ once, by Lemma \ref{eq-oracle-oa}, $O_A$ is constructed by querying $O_{F1}$ $O(c)$ times and $O_{F2}$ $O(1)$ times. By Lemma \ref{init-state}, $|0\rangle|y_{in}\rangle$ can be prepared by querying $O_{u}$ $O(c)$ times.

    Substituting Eq.(\ref{eq-sc}), Eq.(\ref{eq-kappac}) into Eq.(\ref{eq-qlsa-complexity}) and considering the choice of all parameters, the query complexity of solving Eq.(\ref{3-30-2}) for $O_{F1}$, $O_{F2}$ and $O_u$ is
    \begin{equation}
        O\left(sT(\Vert F_1\Vert+\Vert F_2 \Vert){\rm poly}\left(\log\left(\frac{g\eta sT\Vert F_1\Vert\Vert F_2 \Vert}{\epsilon(1-2K^2)\Vert u_{in}\Vert}\right)\right)\right).
    \end{equation}

    Using amplitude amplification algorithm\cite{brassard2002quantum}, we repeat the above process $O(1/\sqrt{p})$ times and get $|u_{out}(T)\rangle=|\bar{x}_{l,0}^{'}(T)\rangle$ which satisfies $\Vert |u_{out}(T)\rangle-u(T)/\Vert u(T)\Vert\Vert \leq \epsilon$ with $\Omega(1)$ success probability.

    The query complexity of the whole process for $O_{F1}$, $O_{F2}$ and $O_u$ is
    \begin{equation}
        O\left(\frac{g\eta s T(\Vert F_1\Vert+\Vert F_2 \Vert)}{\sqrt{1-2K^2}\Vert u_{in}\Vert}{\rm poly}\left(\log\left(\frac{g\eta sT\Vert F_1\Vert\Vert F_2 \Vert}{\epsilon(1-2K^2)\Vert u_{in}\Vert}\right)\right)\right).
    \end{equation}
    The gate complexity of this algorithm is larger than its query complexity by a factor of 
    \begin{equation}
        O\left({\rm poly}\left(\log\left(\frac{ng\eta sT\Vert F_1\Vert\Vert F_2 \Vert}{\epsilon(1-2K^2)\Vert u_{in}\Vert}\right)\right)\right).
    \end{equation}

\end{Prf}

\section{Conclusion and Discussion}\label{sec-discussion}

In this paper, we presented a quantum homotopy perturbation method for solving nonlinear dissipative ODEs. 
The gate complexity of our algorithm is $O(g\eta T{\rm poly}(\log(nT/\epsilon)))$. The complexity of the optimal classical algorithm for solving Eq.(\ref{nonlinear_eq}) is at least linear with $n$, the complexity of the algorithm proposed in \cite{liu2021efficient} is linear with $1/\epsilon$, so our algorithm provides exponential improvement over the best classical algorithms or previous quantum algorithms in $n$ or $\epsilon$. 
$\eta$ and $g$ also affect the complexity of our algorithm, $\eta$ measures the decay of $u(T)$ and increases exponentially as $T$ increases, $g$ measures the decay of $\vec{y}(T)$ defined in Eq.(\ref{eq-linear}) and also increases exponentially as $T$ increases. Our algorithm is effective when $T$ is relatively small which makes $\eta$ and $g$ small enough. $\eta$ and $g$ are also affected by $F_2$, when $T$ is relatively small, the trend of $\eta$ and $g$ increasing exponentially with $T$ may not be obvious due to the influence of $F_2$, this case makes our algorithm perform better.
Our algorithm has the potential to accelerate the solution of various nonlinear equations, and can be applied to nonlinear problems in various fields, such as fluid dynamics, biology, finance, etc, thereby accelerating the research progress of nonlinear science.

Our algorithm only discusses time-independent homogeneous quadratic nonlinear ODEs. When solving time-dependent nonlinear ODEs, the algorithm proposed in \cite{berry2017quantum} is not suitable, an alternative way is to  use the algorithm proposed in \cite{berry2014high} to solve the linear ODEs, then the dependence of complexity on error $\epsilon$ becomes $O(\rm{poly}(1/\epsilon))$. 
Is it possible to optimize the complexity of time-dependent quadratic nonlinear ODEs to $O(\rm poly(\log(1/\epsilon)))$ is an open question.

On the other hand, homotopy analysis method\cite{shijun1998homotopy} and its derivatives\cite{ilhan2021fractional,veeresha2021strong,veeresha2021regarding} are similar to homotopy perturbation method. Whether we can use quantum computing to accelerate the execution process of homotopy analysis method and thus construct a quantum homotopy analysis method is also a question to be investigated further.

Furthermore, how to induce nonlinearity in quantum computing is a basic problem when solving nonlinear equations with quantum algorithm. A common method is producing multiple copies of the original system, some nonlinear quantum algorithms contain copy process\cite{leyton2008quantum,lubasch2020variational,lloyd2020quantum}. In \cite{joseph2020koopman}, a linearization technique of nonlinear classical dynamics based on Koopman-von Neumann method is proposed. \cite{engel2021linear} summarizes three classical linear embedding techniques, including Carleman embedding(Carleman linearization is also called Carleman embedding)\cite{carleman1932application,kowalski1991nonlinear}, coherent states embedding\cite{kowalski1991nonlinear,kowalski1994methods} and position-space embedding\cite{koopman1931hamiltonian}, and then puts forward the prospects of these linear embedding techniques to construct effective quantum algorithms.
An open question is whether there are other ways to induce nonlinearity in quantum computing.

\section*{Acknowledgements}

This work was supported by the National Key Research and Development Program of China (Grant No. 2016YFA0301700), the National Natural Science Foundation of China (Grants Nos. 11625419), the Strategic Priority Research Program of the Chinese Academy of Sciences (Grant No. XDB24030600), and the Anhui Initiative in Quantum Information Technologies (Grants No. AHY080000). 

\section*{Appendix}

In this appendix, we give some lemmas used in proving some conclusions of our work. Lemma \ref{berry-lemma13}, Lemma \ref{berry-lemma14} and Lemma \ref{berry-lemma15} are given in \cite{berry2017quantum}, we just list them again.  

\begin{lemma}\label{lemma-1}
    Let $\gamma \geq 1$, $m\in N^+$, when $t\geq 0$, we have
    \begin{equation}
        \sum_{j=0}^{m-1}{\frac{t^j}{j!}{e^{-\gamma t}}}\leq m.
    \end{equation}
\end{lemma}
\begin{Prf}
    We consider two cases: (1)$0<t\leq m-1$; (2)$t\geq m$.
    
    When $0<t\leq m-1$, we can find $i\in[m-1]_0$ which satisfies $t\in(i,i+1]$, then we have
    \begin{equation}
        \sum_{j=0}^{m-1}{\frac{t^j}{j!}{e^{-\gamma t}}}<m\frac{t^i}{i!}e^{-\gamma t}.
    \end{equation}
    We define
    \begin{equation}
        g(t)=m\frac{t^i}{i!}e^{-\gamma t},\ 0<t\leq m-1.
    \end{equation}
    It is obvious that $g(t)\leq g(\frac{i}{\gamma})$ for $0<t\leq m-1$.
    Using Stirling's formula $i!\approx \sqrt{2\pi i}(\frac{i}{e})^i$, we have
    \begin{equation}
        g(t)\leq m\frac{\left(\frac{i}{\gamma}\right)^i}{\sqrt{2\pi i}(\frac{i}{e})^i}e^{-i}\leq m\left(\frac{1}{\gamma}\right)^{i}\leq m.
    \end{equation}
    Therefore,
    \begin{equation}
        \sum_{j=0}^{m-1}{\frac{t^j}{j!}{e^{-\gamma t}}}\leq m.
    \end{equation}

    When $t\geq m$, we have
    \begin{equation}
        \sum_{j=0}^{m-1}{\frac{t^j}{j!}{e^{-\gamma t}}}<m\frac{t^{m-1}}{(m-1)!}e^{-\gamma t}.
    \end{equation}
    Similar with the case $0<t\leq m-1$, we also have
    \begin{equation}
        m\frac{t^{m-1}}{(m-1)!}e^{-\gamma t}\leq m\frac{\left(\frac{m-1}{\gamma}\right)^{m-1}}{\sqrt{2\pi (m-1)}\left(\frac{m-1}{e}\right)^{m-1}}e^{-(m-1)}\leq m\left(\frac{1}{\gamma}\right)^{m-1}\leq m.
    \end{equation}
    Therefore, for any $t\geq 0$, we have $\sum_{j=0}^{m-1}{\frac{t^j}{j!}{e^{-\gamma t}}}\leq m$.
\end{Prf}

\begin{lemma}\label{lemma-new1}
    Let $\gamma,\beta\in \mathbbm{R}^+$, $m\in N^+$, when $t\geq 0$ and $\gamma/\beta\geq 1$, we have
    \begin{equation}\label{eq-1103-01}
        \sum_{j=0}^{m-1}{\frac{(\beta t)^j}{j!}{e^{-\gamma t}}}\leq m.
    \end{equation}
\end{lemma}
\begin{Prf}
    We define $t^{'}=\beta t$, $\gamma^{'}=\gamma/\beta$, then Eq.(\ref{eq-1103-01}) is written as
    \begin{equation}\label{eq-1103-02}
        \sum_{j=0}^{m-1}{\frac{(t^{'})^j}{j!}{e^{-\gamma^{'} t^{'}}}}\leq m.
    \end{equation}
    By Lemma \ref{lemma-1}, we directly obtain Eq.(\ref{eq-1103-02}).
\end{Prf}

\begin{lemma}\label{lemma-new7}
    Given a matrix $M$ satisfies $\Vert M \Vert \leq 1$ and $\Vert e^{Mt} \Vert\leq \Delta$ for any $t\geq 0$. Let $k,m\in N^{+}$ and satisfy $\frac{2}{(k+1)!}m\Delta(\Delta+1)\leq 1$, $T_k(M)=\sum_{l=0}^{k}{\frac{(Ah)^l}{l!}}$. Then for any $l\in[m]_0$, we have
    \begin{equation}\label{eq0423-1}
        \Vert e^{Ml}-T_k^l(M)\Vert \leq \frac{2l\Delta(\Delta+1)}{(k+1)!}.
    \end{equation}
\end{lemma}
\begin{Prf}
    When $l=0$, $\Vert e^{Ml}-T_k^l(M)\Vert=0$.
    When $l=1$,
    \begin{equation}\label{eq060205}
      \Vert e^{M}-T_k(M)\Vert=\Vert \sum_{j=k+1}^{\infty}{\frac{M^j}{j!}}\Vert\leq \sum_{j=k+1}^{\infty}{\frac{\Vert M\Vert^j}{j!}}\leq \sum_{j=k+1}^{\infty}{\frac{1}{j!}}\leq \frac{2}{(k+1)!}\leq \frac{2}{(k+1)!}\Delta(\Delta+1).
    \end{equation}
    Assuming when $l\leq l^{'}$, we have
    \begin{equation}
        \Vert e^{Ml}-T_k^l(M)\Vert \leq \frac{2}{(k+1)!}l\Delta(\Delta+1),
    \end{equation}
    then by $\Vert e^{Ml} \Vert\leq \Delta$, we have
    \begin{equation}
        \Vert T_k^l(M)\Vert\leq \Delta+\frac{2}{(k+1)!}l\Delta(\Delta+1)\leq \Delta+1.
    \end{equation}
    When $l=l^{'}+1$,
    \begin{align}
        \Vert e^{M(l^{'}+1)}-T_k^{l^{'}+1}(M)\Vert=&\Vert (e^{M}-T_k(M))(\sum_{j=0}^{l^{'}}{e^{Mj}T_k^{l-j}(M)})\Vert\notag\\
        \leq & \frac{2\Delta}{(k+1)!}\times(\sum_{j=0}^{l^{'}}{\Vert T_k^{l-j}(M)\Vert })\notag\\
        \leq & \frac{2}{(k+1)!}(l^{'}+1)\Delta(\Delta+1).
    \end{align}
    Therefore, for any $l\in[m]_0$, we have $\Vert e^{Ml}-T_k^l(M)\Vert \leq \frac{2}{(k+1)!}l\Delta(\Delta+1)$.
\end{Prf}

\begin{lemma}\label{berry-lemma13}
    (\cite{berry2017quantum}). Let $|\psi\rangle$ and $|\varphi\rangle$ be two vectors such that $\Vert |\psi\rangle\Vert \geq \alpha>0$ and $\Vert |\psi\rangle-|\varphi\rangle\Vert \leq \beta$. Then 
    \begin{equation}
        \left\Vert \frac{|\psi\rangle}{\Vert |\psi\rangle \Vert}-\frac{|\varphi\rangle}{\Vert |\varphi\rangle \Vert} \right\Vert \leq \frac{2\beta}{\alpha}.
    \end{equation}
\end{lemma}

\begin{lemma}\label{berry-lemma14}
    (\cite{berry2017quantum}). Let $|\psi\rangle=\alpha|0\rangle|\psi_0\rangle+\sqrt{1-\alpha^2}|1\rangle|\psi_1\rangle$ and $|\varphi\rangle=\beta|0\rangle|\varphi_0\rangle+\sqrt{1-\beta^2}|1\rangle|\varphi_1\rangle$, where $|\psi_0\rangle$, $\psi_1\rangle$, $|\varphi_0\rangle$, $|\varphi_1\rangle$ are unit vectors, and $\alpha,\beta\in [0,1]$. Suppose $\Vert |\psi\rangle-|\varphi\rangle \Vert \leq \delta<\alpha$. Then $\Vert |\psi_0\rangle-|\varphi_0\rangle \Vert \leq \frac{2\delta}{\alpha-\delta}$.
\end{lemma}

\begin{lemma}\label{berry-lemma15}
    (\cite{berry2017quantum}). Let $|\psi\rangle=\alpha|0\rangle|\psi_0\rangle+\sqrt{1-\alpha^2}|1\rangle|\psi_1\rangle$ and $|\varphi\rangle=\beta|0\rangle|\varphi_0\rangle+\sqrt{1-\beta^2}|1\rangle|\varphi_1\rangle$, where $|\psi_0\rangle$, $\psi_1\rangle$, $|\varphi_0\rangle$, $|\varphi_1\rangle$ are unit vectors, and $\alpha,\beta\in [0,1]$. Suppose $\Vert |\psi\rangle-|\varphi\rangle \Vert \leq \delta<\alpha$. Then $\beta\geq \alpha-\delta$.
\end{lemma}

\bibliography{qhpm}

\end{document}